\documentclass[manuscript,screen,review=false]{acmart}

\usepackage{subfigure}
\usepackage{color}
\usepackage{soul}
\usepackage{multicol}
\usepackage{multirow}
\usepackage{tabularx}
\usepackage{booktabs}
\usepackage{xcolor}
\usepackage{xargs}
\usepackage{graphbox}
\usepackage{fontawesome5}
\usepackage{listings}
\usepackage{algorithmic}
\usepackage{tcolorbox}
\usepackage{wrapfig}
\usepackage{enumitem}
\usepackage[normalem]{ulem}\usepackage[colorinlistoftodos,prependcaption,textsize=small]{todonotes}
\usepackage{lineno}

\soulregister\cite7
\soulregister\ref7

\definecolor{googler}{rgb}{0.91, 0.25, 0.21}
\definecolor{googleg}{rgb}{0.45, 0.61, 0.12}
\definecolor{googleb}{rgb}{0.26, 0.52, 0.95}
\definecolor{googley}{rgb}{0.98, 0.73, 0.01}

\definecolor{yesColor}{rgb}{0.565, 0.773, 0.482}
\definecolor{noColor}{rgb}{0.796, 0.447, 0.416}
\definecolor{maybeColor}{rgb}{0.98, 0.73, 0.01}

\newcommand{\YES}{\textcolor{yesColor}{\faCheck}}

\newcommand{\NO}{\textcolor{noColor}{\faTimes}}

\newcommandx{\info}[2][1=inline]{\todo[linecolor=googley, backgroundcolor=googley!20, bordercolor=googley, #1, caption={\textbf{\textcolor{googley}{INFO}}}]{\textcolor{black} {#2} }}
\newcommandx{\help}[2][1=inline]{\todo[linecolor=googleb, backgroundcolor=googleb!20, bordercolor=googleb, #1, caption={\textbf{\textcolor{googleb}{TO DO}}}]{\textcolor{black} {#2} }}

\definecolor{code}{rgb}{0.98, 0.98, 0.98}

\colorlet{punct}{red!60!black}
\definecolor{background}{HTML}{EEEEEE}
\definecolor{delim}{RGB}{20,105,176}
\colorlet{numb}{magenta!60!black}

\lstdefinelanguage{json}{
    basicstyle=\scriptsize\ttfamily,
    numbers=none,
    numberstyle=\scriptsize,
    stepnumber=1,
    numbersep=8pt,
    showstringspaces=false,
    breaklines=true,
    frame=lines,
    backgroundcolor=\color{code},
    literate=
     *{0}{{{\color{numb}0}}}{1}
      {1}{{{\color{numb}1}}}{1}
      {2}{{{\color{numb}2}}}{1}
      {3}{{{\color{numb}3}}}{1}
      {4}{{{\color{numb}4}}}{1}
      {5}{{{\color{numb}5}}}{1}
      {6}{{{\color{numb}6}}}{1}
      {7}{{{\color{numb}7}}}{1}
      {8}{{{\color{numb}8}}}{1}
      {9}{{{\color{numb}9}}}{1}
      {:}{{{\color{punct}{:}}}}{1}
      {,}{{{\color{punct}{,}}}}{1}
      {\{}{{{\color{delim}{\{}}}}{1}
      {\}}{{{\color{delim}{\}}}}}{1}
      {[}{{{\color{delim}{[}}}}{1}
      {]}{{{\color{delim}{]}}}}{1},
}

\AtBeginDocument{%
  \providecommand\BibTeX{{%
    \normalfont B\kern-0.5em{\scshape i\kern-0.25em b}\kern-0.8em\TeX}}}

\acmJournal{CSUR}
\acmYear{2025} \acmVolume{1} \acmNumber{1} \acmArticle{1} \acmMonth{1} \acmPrice{}\acmDOI{10.1145/3722215}

\begin{document}

\title{I/O in Machine Learning Applications on HPC Systems: A 360-degree Survey}

\author{Noah Lewis}
\orcid{0009-0005-1732-7776}
\email{lewis.3621@buckeyemail.osu.edu}
\affiliation{
    \institution{The Ohio State University}
    \city{Baton Rouge}
    \state{Louisiana}
    \country{and}
    \institution{Louisiana State University}
    \city{Columbus}
    \state{Ohio}
    \country{USA}
}
\author{Jean Luca Bez}
\orcid{0000-0002-3915-1135}
\email{jlbez@lbl.gov}
\affiliation{
    \institution{Lawrence Berkeley National Laboratory}
    \city{Berkeley}
    \state{Califorina}
    \country{USA}
}
\author{Suren Byna}
\orcid{0000-0003-3048-3448}
\email{byna.1@osu.edu}
\affiliation{
    \institution{The Ohio State University}
    \city{Columbus}
    \state{Ohio}
    \country{USA}
}


\begin{abstract}
Growing interest in Artificial Intelligence (AI) has resulted in a surge in demand for faster methods of Machine Learning (ML) model training and inference. This demand for speed has prompted the use of high performance computing (HPC) systems that excel in managing distributed workloads. Because data is the main fuel for AI applications, the performance of the storage and I/O subsystem of HPC systems is critical. In the past, HPC applications accessed large portions of data written by simulations or experiments or ingested data for visualizations or analysis tasks. ML workloads perform small reads spread across a large number of random files. This shift of I/O access patterns poses several challenges to modern parallel storage systems. In this paper, we survey I/O in ML applications on HPC systems, and target literature within a $6$-year time window from $2019$ to $2024$. We define the scope of the survey, provide an overview of the common phases of ML, review available profilers and benchmarks, examine the I/O patterns encountered during offline data preparation, training, and inference, and explore I/O optimizations utilized in modern ML frameworks and proposed in recent literature. Lastly, we seek to expose research gaps that could spawn further R\&D.
\end{abstract}



\maketitle

\section{Introduction}
\label{sec:introduction}

Existing High-Performance Computing (HPC) Input and Output (I/O) research has focused on techniques to optimize performance when running traditional HPC application workloads, which typically include simulations and checkpointing partial results~\cite{analyzing-the-io-patterns-of-deep-learning-applications}. Because of the increased popularity of Machine Learning (ML) workloads, there is a rising demand for I/O systems that can effectively accommodate their distinct I/O access patterns. Write operation bursts commonly dominate traditional workloads; however, ML workloads are usually read-intensive and use many small files~\cite{understanding-hpc-application-i-o-behavior-using-system-level-statistics}. Due to the absence of a well-established consensus on the preferred I/O stack for ML workloads, numerous developers resort to crafting their own ad hoc algorithms and storage systems to accommodate the specific requirements of their applications~\cite{understanding-and-leveraging-the-io-patterns-of-emerging-machine-learning-analytics}. This can result in suboptimal application performance due to the under-utilization of the storage system, prompting the necessity for novel I/O optimization methods tailored to the demands of ML workloads.

\begin{figure}[h]
\includegraphics[width=0.6\textwidth]{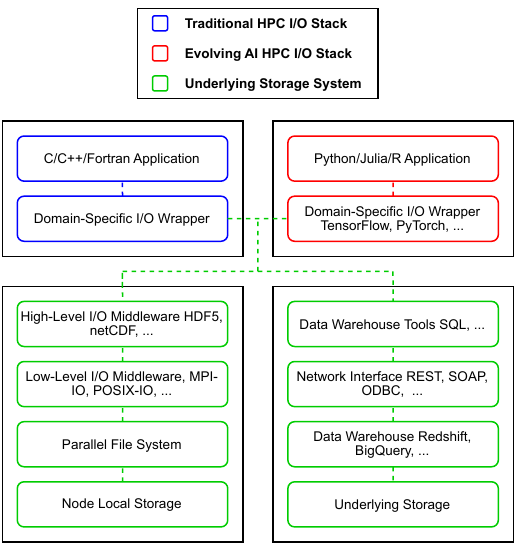}
\caption{Traditional HPC I/O stack vs evolving ML I/O HPC stack (adapted from~\cite{analyzing-the-io-patterns-of-deep-learning-applications}).}
\label{fig:hpc-vs-dl-io-stack}
\centering
\end{figure}

In Fig. \ref{fig:hpc-vs-dl-io-stack}, we show the I/O stack used to run ML workloads (on the right side) compared to the traditional HPC I/O stack (on the left side)~\cite{analyzing-the-io-patterns-of-deep-learning-applications}. The traditional HPC I/O stack has been developed to support massive parallelism. Applications are typically written in C/C++/Fortran languages that use the MPI programming model and I/O middleware, such as HDF5 \cite{hdf5:1997:library, Byna:2020:ExaHDF5:JCST} and netCDF \cite{lee:2008:netcdf}, which internally use MPI-IO and POSIX-IO interfaces to move (read or write) data between main memory and parallel file systems (PFS). In contrast, ML workloads are typically written in productivity-oriented languages, such as Python, and use PyTorch and TensorFlow frameworks. 

Although many ML workloads still use the traditional I/O stack on HPC systems, some are shifting to use remote cloud computing systems that offer services such as data warehouses to store and retrieve large datasets. Cloud platforms such as Google Cloud Platform, AWS, and Microsoft Azure offer access to large datasets, including Google Analytics data~\cite{google-analytics-bigquery} hosted on BigQuery~\cite{google-bigquery}, the $1000$ Genomes Project~\cite{1000-genomes-aws} hosted on Amazon Redshift~\cite{aws-public-datasets}, and NYC Taxi \& Limousine Commission data~\cite{nyc-taxi-azure} hosted on Azure Synapse Analytics~\cite{azure-synapse}. HPC workloads that aim to use these datasets require near-first-class integration with the cloud to enable efficient data transfers to HPC systems. When using the traditional I/O stack, high bandwidth and low latency data access are possible due to the locality of the data and the highly optimized stack. In contrast, access to data stored in a data warehouse is often accomplished through a network interface such as REST~\cite{rest-interface-architecture-styles-and-the-design-of-network-based-software-architectures}, SOAP~\cite{soap-interface}, or ODBC~\cite{odbc-interface}. Given the high latency inherent in accessing data stored in a remote storage location, optimizing data retrieval is crucial for reasonable performance. Despite the lower I/O performance imposed when using a data warehouse, they are still preferred in many scenarios because they provide a centralized data storage location, facilitate data integration from various sources, and enable data governance, which can result in higher-quality data. High-quality data is essential when training ML models because it directly influences their accuracy, reliability, and generalizability~\cite{overview-and-importance-of-data-quality-for-machine-learning-tasks}. In addition, the monetary cost needed to purchase and maintain an HPC system can far outweigh the cost of using managed cloud services~\cite{cost-effective-hpc-the-community-or-the-cloud, a-comparative-study-of-high-performance-computing-on-cloud}. Furthermore, integration of HPC systems with available cloud resources is necessary due to the cloud being used as a data storage solution from which datasets can be publicly shared and used by all.

The differences listed above in the I/O stacks for ML applications using the traditional PFS compared to cloud computing systems motivate the development of novel storage system designs and I/O performance optimization techniques to adapt to the I/O operations commonly found in ML processes. Ideally, the methods would be transparent to the application developers. Notably, many scientists may lack formal training in software development, as evidenced by Paul et al.~\cite{characterizing-machine-learning-i-o-workloads-on-leadership-scale-hpc-systems}. The study revealed that ML workloads are becoming ever more popular and that developers from various scientific domains beyond computer science often overlook the utilization of node-local storage (i.e., burst buffers) within HPC systems. Further evidence for the increasing interest in distributed training is the number of posts on the popular tech website StackOverflow. The cumulative number of posts discussing distributed training increased staggeringly from $257$ posts in $2015$ to $2,061$ posts in 2024 \cite{stackoverflow_distributed_training}. The increasing number of ML workloads that are running on HPC systems prompts the need for transparent I/O optimization techniques to decrease the complexity required for scientists to write software that takes advantage of the resources offered by HPC systems.

Furthermore, scientists, I/O performance enthusiasts, and HPC practitioners can gain significant insights from benchmarks that accurately emulate ML workloads and profilers that capture critical I/O information from the underlying storage systems. Using these tools, individuals can find potential I/O bottlenecks within their ML applications. Access to realistic benchmarking tools and fine-grained profilers is also invaluable for HPC storage research, allowing researchers to analyze and emulate the diverse range of I/O operations found in ML workloads. Currently, I/O analysis tools often fall short in a number of ways, including the inability to identify I/O requests at the thread level and only capturing traces of specific I/O interfaces (i.e., POSIX and STDIO). This makes it challenging to find a general-purpose tool applicable to a wide variety of workloads, which can reduce the consistency in which I/O traces are captured. In addition, many tools (i.e., Darshan~\cite{modular-hpc-i-o-characterization-with-darshan} and Scalasca~\cite{geimer:2010:scalasca}) target MPI applications limiting their compatibility and effectiveness with other communication libraries (i.e., GLOO~\cite{gloo-github}, Spark~\cite{spark-docs}, and Dask~\cite{dask-docs}). It should be noted that Darshan is capable of capturing I/O traces from applications that do not use MPI through the \texttt{NONMPI} environment variable. However, this results in the output of a single DXT log for each process instead of the more common single file output that captures all I/O traces from an application using MPI. In some cases, this leaves developers with the task of manually merging the results from their non-MPI applications in order to conduct further analysis.

The topic of ML I/O and whether it has a significant role in poor performance has been a debate within the HPC community. Many point out that HPC systems often have enough node-local storage to cache the dataset, resulting in minimal time spent waiting on I/O~\cite{streamlining-distributed-deep-learning-io-with-ad-hoc-file-systems, ml-perf-hpc, characterizing-machine-learning-i-o-workloads-on-leadership-scale-hpc-systems, zhu:2018:entropy-aware}. Others mention that not all HPC systems have large amounts of node local storage or the datasets used are simply too large to be cached~\cite{streamlining-distributed-deep-learning-io-with-ad-hoc-file-systems, ml-perf-hpc, data-aware-storage-tiering-for-deep-learning, zhu:2018:entropy-aware}. In addition, due to the random sampling that commonly occurs during ML training, modern PFSs can often become an I/O bottleneck~\cite{inside-the-lustre-file-system, io-performance-characterization-and-prediction-through-machine-learning-on-hpc-systems, zhu:2018:entropy-aware}. Ultimately, whether I/O is a significant issue is subjective to the workload at hand. It depends on various factors such as dataset size, storage capacity, storage system bandwidth, I/O library implementation and their configurations, and I/O access patterns of applications. Furthermore, many ML application developers are not HPC storage experts; hence, even if their system is capable of providing great performance, transparent solutions such as caching and prefetching are needed to ensure the HPC systems are utilized effectively. Transparent solutions are of particular interest due to the increasing number of heterogeneous processors used within ML applications. These increase the complexity of optimally utilizing available compute resources. For these reasons, we believe that I/O and data management are key areas of interest to the ML community.

There are several common phases in an ML application lifecycle: \textit{data generation}, \textit{data preparation}, \textit{training}, and \textit{inference}. The data management found within each phase varies widely due to different I/O access patterns, data sources, and file formats. We provide a taxonomy of the data management found throughout the ML lifecycle in Fig.~\ref{fig:radial-tree}. The ML lifecycle begins with data generation, which involves collecting data from various sources (e.g., simulation results or web scraping), which are then stored using a chosen file format (e.g., .tfrecord, .csv, or .png) and storage location (e.g., SSD, disk, cloud, or database). This is commonly followed by an offline data preparation phase, which performs a series of transformations and reductions (i.e., filters, aggregations, or generalizations) over the collected data to increase the training efficiency and accuracy of the ML model. The training process involves reading samples from the data sources, which are then fed to computations. The order in which the samples are read largely depends on the chosen optimizer (e.g., Stochastic Gradient Descent and Adaptive Moment Estimation or Adam, etc.) and training distribution strategy (e.g., model or data parallelism). Due to many optimizers requiring random sampling and storage systems being optimized for contiguous sequential reading, training I/O is a potential bottleneck. Once the model has been trained, it is often deployed and fed ``new'' data during the inference phase. It is common for the online and offline data preparation phases to occur before and during the inference phase as well. More details on each phase of ML can be found in Section~\ref{sec:ml-phases}, but it is important to note the wide range of factors that impact I/O and data management throughout the ML lifecycle, such as varying file formats, data sources, data preparation requirements, and data modalities, among others. It is impractical to cover all possible combinations of the factors mentioned above, and therefore, in this paper, we aim to discuss the I/O approaches found in the most common ML applications running on HPC systems.

The structure of the paper is as follows. First, in Section~\ref{sec:scope-of-the-survey}, we define the scope of the survey and present trends in the cited literature. Then, in Section~\ref{sec:ml-phases}, we describe the data formats and modalities found in ML workloads and provide an overview of the typical phases of ML applications and the popular methods of distributing the model training on HPC systems. Next, in Section~\ref{sec:ml-io-profiling-and-benchmarks}, we describe the benchmarks and profiling tools scientists and researchers can use to ensure proper system utilization under ML workloads. In Section~\ref{sec:ml-io-access-patterns}, we describe the I/O access patterns found in the offline data preparation, training phase, and inference phases. The training phase is exemplified by analyzing the currently available representative I/O benchmarks. We then discuss, in Section~\ref{sec:ml-io-optimization-techniques}, various I/O optimization techniques used in modern ML frameworks and those proposed in the recent research literature. Lastly, in Section~\ref{sec:gaps-in-ml-io-research}, we discuss the gaps in the current research of ML I/O in HPC systems.

\section{Scope of the Survey}
\label{sec:scope-of-the-survey}

ML applications use various environments, from local machines for small-scale applications to HPC and cloud systems for large-scale modeling. This survey aims to provide a comprehensive review of the I/O patterns in ML running on HPC systems. This means that I/O patterns and other data management solutions that may be common in other environments (e.g., cloud or local) are not relevant. For example, in Sec~\ref{sec:ml-io-access-patterns}, we discuss only the I/O access patterns found in common HPC scenarios and exclude I/O access patterns that may be common in cloud or local environments.

\begin{table}[ht]
\centering
\begin{tabular}{cccccccccccccl}
\rotatebox{90}{GPU Mem. Hierarchy} & \rotatebox{90}{File Format} & \rotatebox{90}{Sample Caching} & \rotatebox{90}{Sample Prefetching} & \rotatebox{90}{Sample Shuffling} &
\rotatebox{90}{Asynchronous Sampling} & \rotatebox{90}{Profiling} &
\rotatebox{90}{Benchmarking} & \rotatebox{90}{Model Checkpointing} &
\rotatebox{90}{Model Storage Offloading} & \rotatebox{90}{Scheduling} &
\rotatebox{90}{ML Analysis} & \rotatebox{90}{apers}  & \textbf{References} \\
\toprule
\NO & \NO  & \NO 
& \NO  & \NO 
& \NO  & \NO 
& \YES & \NO 
& \NO  & \NO 
& \NO  & $6$ & \cite{open-graph-benchmark, storage-benchmarking-with-deep-learning-workloads, dlio-a-data-centric-benchmark-for-scientific-deep-learning-applications, profiling-and-improving-the-pytorch-dataloader-for-high-latency-storage, ibench, ibench-evaluation} \\
\NO & \NO & \NO 
& \NO  & \NO 
& \NO  & \NO 
& \NO  & \NO 
& \NO  & \NO 
& \YES & $5$ & \cite{characterizing-machine-learning-i-o-workloads-on-leadership-scale-hpc-systems, analyzing-the-io-patterns-of-deep-learning-applications, a-case-study-of-data-management-challenges-presented-in-large-scale-machine-learning-workflows, high-performance-io-for-large-scale-deep-learning, analyzing-the-distributed-training-of-deep-learning-models-via-data-locality} \\
\NO & \NO  & \NO  
& \NO  & \NO  
& \NO  & \NO 
& \NO  & \YES 
& \NO  & \NO 
& \NO  & $5$ & \cite{evaluating-multi-level-checkpointing-for-distributed-deep-neural-network-training, check-n-run-a-checkpointing-for-training-deep-learning-recommendation-models, deepfreeze-towards-scalable-asynchronous-checkpointing-of-deep-learning-models, a-study-of-checkpointing-in-large-scale-training-of-deep-neural-networks, datastates-llm} \\
\NO & \NO  & \YES 
& \NO  & \NO  
& \NO  & \NO 
& \NO  & \NO  
& \NO  & \NO 
& \NO  & $4$ & \cite{icache, analyzing-data-reference-characteristics-of-deep-learning-workloads-for-improving-buffer-cache-performance, a-deep-learning-dataloader-with-shared-data-preparation, mmdataloader} \\
\NO & \NO  & \YES
& \YES & \NO 
& \NO  & \NO 
& \NO  & \NO 
& \NO  & \NO 
& \NO  & $3$ & \cite{clairvoyant-prefetching-for-distributed-machine-learning-i-o, monarch, high-performance-io-for-large-scale-deep-learning} \\
\NO & \NO  & \NO 
& \NO  & \NO 
& \NO  & \NO 
& \NO  & \YES
& \YES & \NO 
& \NO  & $3$ & \cite{smart-infinity-fast-large-language-mode-training-using-near-storage-processing-on-a-real-system, zero-infinity-breaking-the-GPU-memory-wall-for-extreme-scale-deep-learning, stronghold-fast-and-affordable-billion-scale-deep-learning-model-training} \\
\NO & \NO  & \NO 
& \NO  & \NO 
& \NO  & \NO 
& \NO  & \NO 
& \NO  & \YES
& \NO  & $3$ & \cite{io-performance-characterization-and-prediction-through-machine-learning-on-hpc-systems, adpatively-periodic-io-scheduling-for-concurrent-hpc-applications, deep-lake-a-lakehouse-for-deep-learning} \\
\NO & \YES & \YES
& \YES & \YES
& \NO  & \NO 
& \NO  & \NO 
& \NO  & \NO 
& \NO  & $2$ & \cite{ffcv, diesel-plus-accelerating-deep-learning-tasks-on-image-datasets} \\
\NO & \NO  & \YES
& \NO  & \YES
& \NO  & \NO 
& \NO  & \NO 
& \NO  & \NO 
& \NO  & $2$ & \cite{asynchronous-io-strategy-for-large-scale-deep-learning-applications, efficient-data-loading-for-deep-neural-network-training} \\
\NO & \NO  & \YES
& \NO  & \YES
& \YES & \NO 
& \NO  & \NO 
& \NO  & \NO 
& \NO  & $2$ & \cite{ddstore-distributed-data-store-for-scalable-training-of-graph-neural-networks, accelerating-machine-learning-io-by-overlapping-data-staging-and-mini-batch-generations} \\
\NO & \NO  & \NO  
& \NO  & \NO  
& \NO  & \YES
& \NO  & \NO  
& \NO  & \NO 
& \NO  & $1$ & \cite{tf-darshan-understanding-fine-grained-i-o-performance-in-machine-learning-workloads} \\
\NO & \YES & \NO 
& \NO  & \NO 
& \NO  & \NO 
& \NO  & \NO 
& \NO  & \NO 
& \NO  & $1$ & \cite{progressive-compressed-records-taking-a-byte-out-of-deep-learning-data} \\
\NO & \NO  & \NO 
& \NO  & \YES
& \NO  & \NO 
& \NO  & \NO 
& \NO  & \NO 
& \NO  & $1$ & \cite{why-globally-re-shuffle-revisiting-data-shuffling-in-large-scale-deep-learning} \\
\YES & \NO  & \NO 
& \NO  & \NO
& \NO  & \NO 
& \NO  & \NO 
& \NO  & \NO 
& \NO  & $1$ & \cite{flash-attention-fast-and-memory-efficient-exact-attention-io-awareness} \\
\bottomrule
\end{tabular}
\caption{I/O-related topics discussed in the surveyed ML papers.}
\label{tab:subjects-discussed-in-papers-surveyed}
\end{table}

To ensure a comprehensive review of the current literature, we have selected a $6$-year time window from $2019$ to $2024$ and used ACM Digital Library and IEEE Xplore to filter by papers that mention the following keywords: ``I/O Access Patterns'', ``I/O pattern'', ``HPC'', ``Storage performance'', and ``I/O performance in AI/ML applications''. We found additional papers through the citations of the work initially discovered using the digital libraries for a total of $39$ papers. The various subjects explored in the papers surveyed are shown in Table~\ref{tab:subjects-discussed-in-papers-surveyed}. We selected papers from leading conferences and journals in HPC and AI areas including:

\begin{itemize}[label=--]
    \item The IEEE/ACM Int. Conference for High-Performance Computing, Networking, Storage and Analysis (SC)
    \item The IEEE/ACM International Symposium on Cluster, Cloud and Internet Computing (CCGRID)
    \item ACM Very Large Database Conference (VLDB)
    \item ACM Conference on Cloud Computing, Big Data \& Emerging Topics (JCC-BD\&ET)
    \item ACM Transactions on Parallel Computing (TOPC)
    \item IEEE International Symposium on High-Performance Computer Architecture (HPCA)
    \item IEEE International Conference on Cluster Computing (CLUSTER)
    \item IEEE International Symposium on High-Performance Computer Architecture (HPCA)
    \item IEEE Computer Society Conference on Computer Vision and Pattern Recognition (CVPR)
    \item IEEE International Conference on High Performance Computing, Data, and Analytics (HiPC)
    \item IEEE Conference on High Performance Extreme Computing (HPEC)
    \item IEEE International Parallel and Distributed Processing Symposium (IPDPS)
    \item IEEE Modeling, Analysis, and Simulation of Computer and Telecommunication Systems (MASCOTS)
    \item IEEE International Conference on High Performance Computing \& Simulation (HPCS)
    \item IEEE Transactions on Parallel and Distributed Systems (TPDS)
    \item IEEE Euromicro International Conference on Parallel, Distributed and Network-Based Processing (PDP)
    \item USENIX Networked Systems Design and Implementation (NSDI)
    \item NIPS International Conference on Neural Information Processing Systems (NeurIPS)
    \item SpringerOpen Journal of Big Data
\end{itemize}

\begin{figure}[ht]
\includegraphics[width=13cm]{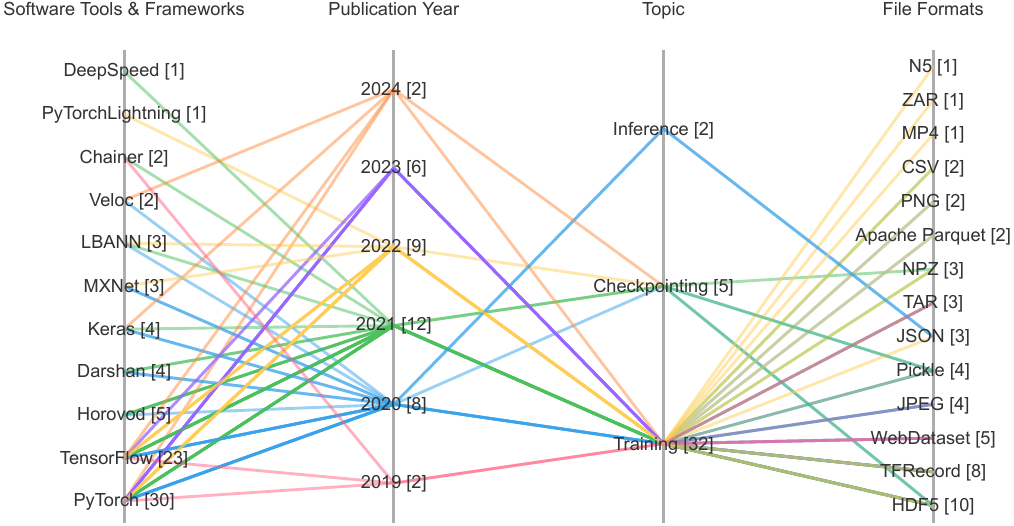}
\caption{Trends in cited literature. Each color represents a year. The number to the right of each label represents the total number of papers associated with that label. The publication year and topic each have $1$ paper. These two categories each sum to the total number of papers $39$. The leftmost and rightmost categories sum to more than $39$ (a paper can discuss more than one item).}
\label{fig:trends-in-ml-io}
\centering
\end{figure}

The parallel coordinates diagram in Fig.~\ref{fig:trends-in-ml-io} illustrates the popularity of various software tools, frameworks, topics, and file formats in the surveyed papers. TensorFlow was discussed in $23$ papers, while PyTorch appeared in $30$ papers, making these two ML frameworks the most popular by a significant margin. Training emerged as the most common area of focus, with a total of $32$ papers, followed by checkpointing, with a total of $5$ papers, and lastly, inference, with $2$ papers. Furthermore, HDF5 and TFRecord were notably mentioned in $8$ and $10$ of the surveyed papers, respectively. Based on these results, there has been a strong focus on optimizing I/O for training ML models using TensorFlow and PyTorch over the last couple of years.

In this paper, our contributions are a presentation and analysis of the I/O patterns found when emulating two renowned ML workloads: BERT~\cite{bert-pre-training-of-deep-bidirectional-transformers-for-language-understanding}, and Unet3D~\cite{unet3d}. In addition, we conduct a review of the currently proposed I/O optimization techniques through R\&D efforts and present several areas where future research is needed. Understanding the I/O patterns found in ML applications allows model training speeds to be reduced. This contributes to the rapid development of ML models, which can be used to perform tasks such as medical imaging analysis and help in the creation of autonomous vehicles. It should be noted that throughout the survey, we use the terms ``AI'' and ``ML'' synonymously, as is commonly done in the industry. However, ``ML'' is the more precise term referring to the concept in which machines extract knowledge from data and learn from it.

\begin{tcolorbox}[colback=white!5!white,colframe=black!75!black,title=Summary $\#1$]
The analysis of the surveyed papers reveals a significant trend towards optimizing I/O during the training phase. Additionally, TensorFlow and PyTorch were by far the most popular ML frameworks. Lastly, the HDF5 and TFRecord file formats were the most popular.
\end{tcolorbox}

\section{ML Data: Data Formats, Modalities, and Common Phases}
\label{sec:ml-phases}
In this section, we describe the ML lifecycle and the data formats and modalities commonly found throughout this lifecycle. We also discuss the common phases ML applications undergo and the ways of distributing these phases to exploit the resources offered by HPC systems. We then discuss model checkpointing and its impact on training speeds.

\begin{figure}[ht]
\includegraphics[width=1\columnwidth]{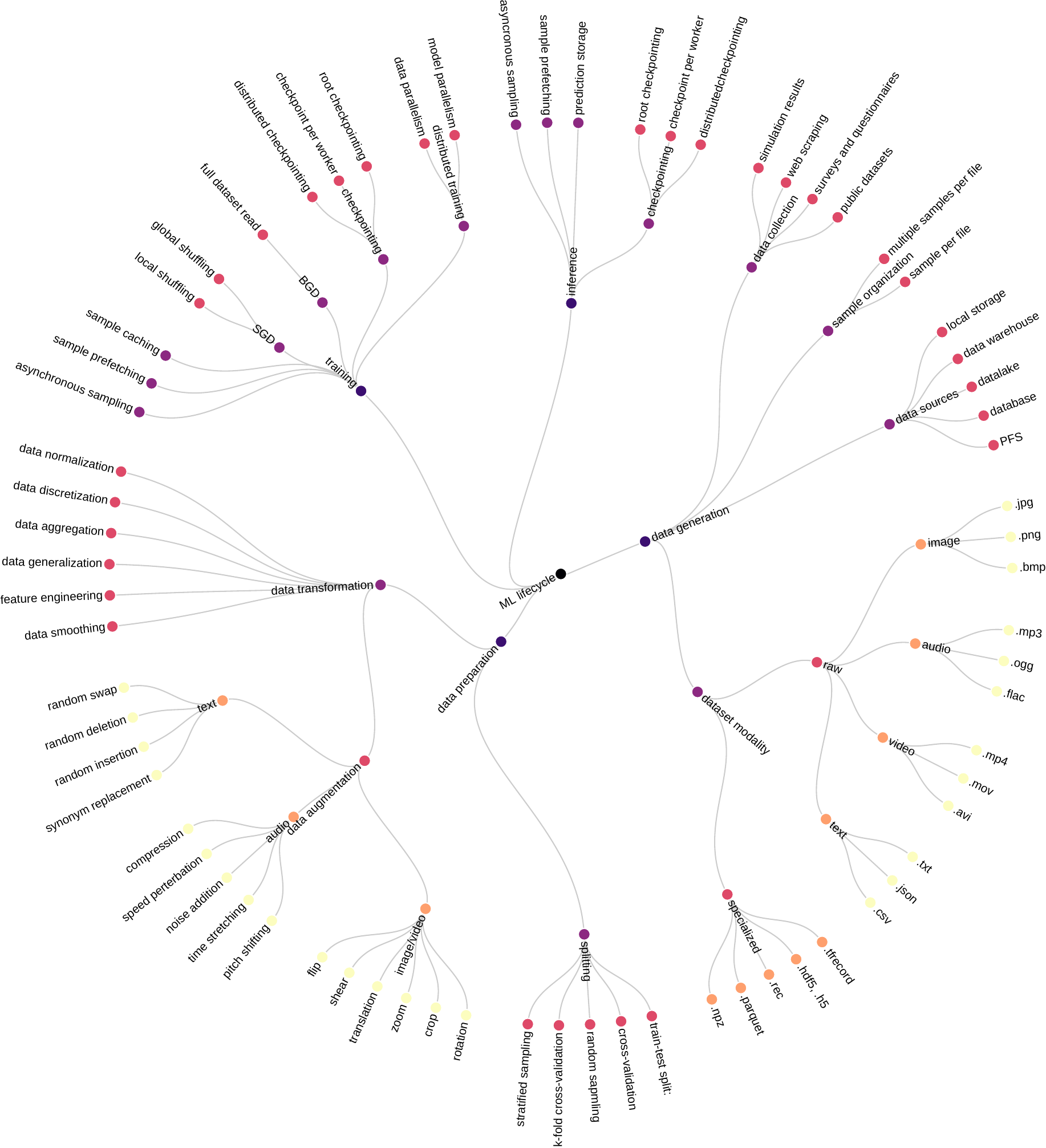}
    \caption{A taxonomy of data management during the ML lifecycle: Data Generation, Dataset Preparation, Training, and Inference. In the data generation phase, we show the different ways data is collected and the various data/file models used in application domains. In the dataset preparation phase, we categorize various operations performed on the data to improve its quality. For the training and inference phases, we categorize commonly used data access patterns and I/O optimizations.}
    \label{fig:radial-tree}
    \centering
\end{figure}

ML applications typically undergo four phases during the ML lifecycle: \textbf{data generation}, \textbf{dataset preparation}, \textbf{training}, and \textbf{inference}~\cite{analyzing-the-io-patterns-of-deep-learning-applications}. We show a taxonomy of the data management operations and characteristics throughout this lifecycle in Fig.~\ref{fig:radial-tree}. During data generation, data is obtained through various sources and methods such as simulations, web scraping, and public datasets. During this phase, properties such as the dataset modality, file format, and data storage location are often decided. The data preparation phase is divided into various data preparation steps, including transformation, augmentation, and splitting. The training arm of the taxonomy consists of data I/O operations across common training algorithms and checkpointing tasks. Inference includes a taxonomy similar to that of the training phase. As can be seen, data is a prominent component throughout the entirety of the ML lifecycle.

\subsection{Data Formats and Dataset Modalities}

Data is at the forefront of the ML lifecycle, and without the correct quantity and quality of data, machine learning models fail to achieve optimal performance and generalization, disabling their ability to provide accurate insights and solutions. Here, we describe the file formats and data modalities commonly used during model training and inference. The file formats described were chosen due to their frequent appearance in the cited literature or their connection with state-of-the-art ML training and preprocessing frameworks (e.g., \textit{.tfrecord} -- TensorFlow and \textit{.parquet} -- Apache Parquet).

\begin{table}[htbp]
    \centering
    \begin{tabularx}{0.95\linewidth}{l*{5}{>{\centering\arraybackslash}X}l}
        \toprule
        \multirow{2}{*}{\textbf{File Format}} & \multicolumn{6}{c}{\textbf{Features}} \\
        \cmidrule(lr){2-7}
        & Parallel I/O & Chunking & Partial \mbox{Compression} & Schema & Binary & File Extension \\
        \midrule
        TFRecord \cite{tensorflow_tfrecord} & \NO & \NO & \NO & \YES$^{1}$ & \YES & .tfrecord \\
        HDF5 \cite{hdf5:1997:library} & \YES & \YES & \YES & \YES & \YES & .h5 .hdf5 $^{2}$  \\
        Apache Parquet \cite{apache-parquet} & \YES  & \YES & \YES & \YES & \YES & .parquet  \\
        RecordIO \cite{recordio-format} & \NO & \YES & \YES & \NO & \YES & .rec \\
        NPZ \cite{numpy} & \NO & \NO & \NO & \NO & \YES & .npz \\
        \midrule
        Image Formats & \NO & \NO & \NO & \NO & \YES & .jpg .png \\
        Video Formats & \NO & \NO & \NO & \NO & \YES & .mp4 .avi \\
        Audio Formats & \NO & \NO & \NO & \NO & \YES & .mp3 .wav\\
        Text Formats & \NO & \NO & \NO & \NO & \NO & .txt .csv \\
        \bottomrule
    \end{tabularx}
    \vspace{5pt}
    \caption{A Comparison of commonly used file formats in AI. The partial decompression property indicates that the file format library transparently performs the decompression operation, possibly after it is enabled or disabled. ($^{1}$TFRecords are read and written using Protocol Buffers (protobufs), which require a schema. $^{2}$While .h5 and .hdf5 are used commonly, HDF5 allows any file extension.)}
    \label{tab:file_formats}
\end{table}

There are many file formats, such as those seen in Table~\ref{tab:file_formats}, which have diverse properties that can affect performance when storing and retrieving ML datasets. While formats such as HDF5~\cite{hdf5:1997:library}, Apache Parquet~\cite{apache-parquet}, and RecordIO~\cite{recordio-format} have been created specifically to target efficient distributed data storing and loading, many applications still rely on unoptimized solutions such as raw text, video, audio, and image formats. These formats are often used due to their ease of use and compatibility with common system software such as standard text editors, video players, audio players, and image viewers. However, in addition to their poor performance, these formats lack schema enforcement.

Schema enforcement gives the developer the ability to confine the domain of the file contents. This commonly ensures attempting to write data outside the input domain will fail. Without it, the risk to data integrity can lead to issues such as hard-to-find bugs caused by data corruption. Compression is a common technique used to reduce the size of data by encoding the information with fewer bits. Partial compression enables data loaders to read and write to subsets of a compressed file without having to load the file in its entirety. Due to ML applications commonly reading small random subsets of the data, the inability to read subsets of data without loading the entire file could create a bottleneck. Chunking refers to the ability to divide the data in the file into discrete, manageable ``chunks.'' 

Apache Parquet and HDF5 enable a high level of parallelism by allowing the user to specify the unit (i.e., row group size and data page size for parquet and chunk size for HDF5) in which the data is stored. In addition, Apache Parquet stores data in a columnar format, allowing the file to be split into row groups and processed independently. HDF5, on the other hand, supports features such as chunking and collective I/O, which enables efficient parallel file processing. In many cases, such as when the data is stored in a Lustre PFS~\cite{sun:2007:lustre, wang:2009:lustre-internals}, this enables parallel reads and writes across independent sections of the file system (e.g., using file striping). Formats such as RecordIO~\cite{recordio-format}, NPZ~\cite{numpy}, and TFRecord~\cite{tensorflow_tfrecord} store data in a more serialized manner. Due to the variety of possible optimizations offered by distributed file formats, making the appropriate selection is essential for optimizing performance on HPC systems.

\begin{table}[!b]
    \centering
    \begin{tabular}{llrrl}
        \toprule
        {\textbf{Modality}} & {\textbf{Dataset}} & {\textbf{Total Files}} & {\textbf{Total Size}} & {\textbf{Used File Formats}} \\
        \midrule
        \multirow{3}{*}{Image} & MNIST~\cite{mnist-dataset} & $4$ & $101.16$ MiB & Custom \\
        & CIFAR-100~\cite{cifar-dataset} & $60,000$ & $184$ MiB & Custom \\
        & ImageNet~\cite{imagenet-a-large-scale-heirarchical-image-database} & $1,431,167$ & $150$ GiB & .jpg, .txt \\
        \midrule
        \multirow{3}{*}{Audio} & UrbanSound8K~\cite{urbansound-dataset} & $18733$ & $6.60$ GiB & .wav, .csv \\
        & LibriSpeech~\cite{librispeech} & $10,000$ & $60$ GiB & .flac, .txt \\
        & AudioSet~\cite{audioset-dataset} & $2,084,320$ & $2.4$ GiB & .tfrecord, .csv \\
        \midrule
        \multirow{3}{*}{Video} & Charades~\cite{charades-dataset} & $9,900$ & $55$ GiB & .csv, .txt, .mat, .jpg Frames \\
        & UCF101~\cite{ufc101} & $13,320$ & $6.69$ GiB & .avi \\
        & HMDB~\cite{hmdb-dataset} & $7,000$ & $2$ GiB & .txt, .avi  \\
        \midrule
        \multirow{3}{*}{Time Series} & NAB~\cite{nab-dataset} & $58$ & $9$ MiB & .csv \\
        & M4~\cite{m4-dataset} & $13$ & $310$ MiB & .csv \\
        & UCR Time Series Archive~\cite{ucr-dataset} & $416$ & $853$ MiB & .tsv \\
        \midrule
        \multirow{3}{*}{Graph} & Cora~\cite{cora-dataset} & $1$ & $270$ KiB & .csv \\
        & Protein-Protein Int.~\cite{protein-dataset} & $1$ & $3,661$ KiB & .csv \\
        & Amazon Co-purchasing~\cite{amazon-dataset} & $1$ & $954,597$ KiB & .txt \\
        \midrule
        \multirow{3}{*}{Text} & refinedweb~\cite{refined-web-dataset} & $2,858$  & $2.8$ TiB & .parquet \\
        & The Pile~\cite{the-pile-dateset} & $17,103,059$ & $825$ GiB & .txt \\
        & The Stack v2~\cite{the-stack-v2-dataset} & $3,028,000,000$ & $67.5$ TiB & Code Files \\
        \bottomrule
    \end{tabular}
    \vspace{5pt}
    \caption{Comparison of different ML dataset modalities.}
    \label{tab:ml_dataset_comparison}
\end{table}

\begin{figure}{}{}
    \includegraphics[width=1\columnwidth]{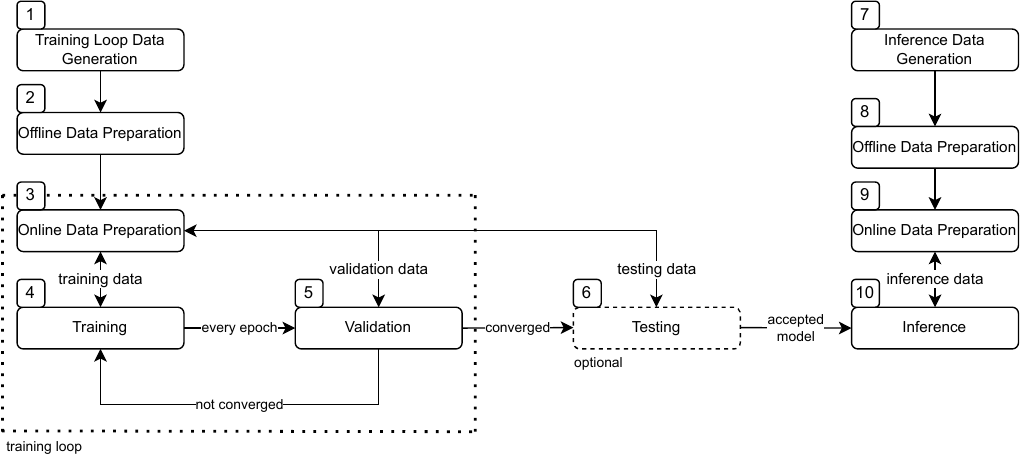}
    \caption{Common phases during the ML lifecycle. The numbers in the top-left corners of each phase indicate their chronological order. Note that while phases $1$ and $2$ are sequential, phases $1$, $2$, and $7$, $8$ can occur in parallel.}
    \label{fig:ml-training-phase-diagram}
    \centering
\end{figure}

The selection of file formats model developers can choose from is often constrained by the data modality in use. The dataset modality refers to the mode or type of data being stored (e.g., audio, video, image, etc.). Due to the large variety of ML model goals (e.g., image recognition, optical character recognition (OCR), relationship detection, etc.), modalities vary widely. The dataset modality has a significant impact on the viable file formats available and, therefore, the common I/O operations required to load and use the dataset. In Table~\ref{tab:ml_dataset_comparison}, we list the dataset modalities commonly used for ML applications and three corresponding publicly available datasets often used for ML research, exemplifying each modality. The provided values are approximate and subject to variation depending on which version of the dataset is selected. For instance, the Charades~\cite{charades-dataset} dataset is available at different video resolutions, while the CIFAR-10~\cite{cifar-dataset} dataset is available as a custom binary format or as a binary containing a Python ``Pickled'' object.  Additionally, in certain cases, like AudioSet, links to the video data are provided rather than the raw video data itself, allowing users to choose the preferred video format.

\subsection{Common ML Phases}
\label{subsection:common-ml-phases}

The \textbf{dataset preparation} phase is critical in transforming the data into a form that is suitable for model training and inference. It is a critical step in the ML lifecycle and is commonly performed before both the training and inference phases. \textbf{Offline dataset preparation} is preprocessing that can be done before the model training or inference phase begins. In contrast, \textbf{online dataset preparation} is preprocessing that occurs in real-time during the inference or training phase. There are several major data transformation techniques commonly used during both offline and online dataset preparation to aid in model training and inference. These are listed below:

\begin{itemize}[label=--]
    \item \textbf{Data Smoothing} is a method used to remove noise from the dataset, ultimately allowing models to identify patterns in the dataset more accurately~\cite{data-smoothing}.
    \item \textbf{Feature Engineering} is used to construct new features from existing attributes in the dataset~\cite{feature-engineering}.
    \item \textbf{Data Generalization} defines hierarchies within the features of the dataset, allowing the data to be analyzed at varying levels of granularity or abstraction~\cite{data-generalization}.
    \item \textbf{Data Aggregation} is used to summarize or reduce features from the dataset to produce meaningful statistics such as mean, median, or sum among many others~\cite{data-aggregation}.
    \item \textbf{Data Discretization} is used to divide continuous values into discrete ones, making analysis easier and improving the efficiency of many algorithms~\cite{data-discretization}.
    \item \textbf{Data Normalization} is used to scale the data changing absolute values while preserving the relative proportions or ratios between data points~\cite{data-normilization}.
    \item \textbf{Data Augmentation} is a strategy used to generate new data from existing data~\cite{understanding-and-leveraging-the-io-patterns-of-emerging-machine-learning-analytics}.
\end{itemize}

Augmenting the dataset allows for training models on a larger dataset, which can lead to better performance. The augmentation methods employed often depend on the dataset modality. For text datasets, examples of augmentations include random swap, random insertion, random deletion, and random synonym replacement~\cite{text-data-augmentation}. In contrast, for image datasets, augmentations include flipping, cropping, and rotation~\cite{image-data-augmentation}. The data transformation techniques used affect both data preprocessing speeds and the amount of data needing to move from the preprocessing phase to the training phase. The variety of data transformation techniques that may need to be applied to the dataset causes the file format to be significant. It has a large impact on data transformation speeds as it affects the time needed to apply a given algorithm to the dataset, e.g., when performing data aggregation where the dataset is filtered based on a feature value, processing data in a columnar format (instead of row-based) can be more efficient due to the data locality~\cite{column-stores-vs-row-stores-how-different-are-they-really}. 

The \textbf{training} phase that occurs during the ML lifecycle can further be categorized, as seen in Fig.~\ref{fig:ml-training-phase-diagram}, into a \textbf{training} phase and a \textbf{validation} phase. The figure shows the training loop and the inference phase. The training loop is often run iteratively for a certain number of epochs until the desired accuracy is achieved. An epoch is defined as one complete pass through the entire dataset~\cite{batch-vs-epoch}. It is the responsibility of the model designers to choose the number of epochs that strikes a balance between enabling the model to find meaningful relationships between the input and target datasets while avoiding excessive epochs that lead to over-fitting. Over-fitting occurs when the model becomes excessively entwined with the training data, prioritizing memorization over the ability to generalize and make accurate predictions on new, unseen data. Furthermore, because the entire dataset is read multiple times, model designers need to select the proper data format for their dataset that allows efficient data retrieval.

There are three datasets known as the training, validation, and testing datasets~\cite{algorithmic-splitting-a-method-for-dataset-preparation} used during the \textbf{training}, \textbf{validation}, \textbf{testing} phases. Although each dataset possesses distinct content, they commonly use the same modality and file formats. All three datasets often require preparation (i.e., cleaning, augmentation, etc.) before they can be used in later stages of the ML lifecycle. Dataset splitting involves dividing the dataset into training, validation, and test sets. Common techniques for this partitioning include random sampling~\cite{scalable-simple-random-sampling-and-stratified-sampling}, \textit{k}-fold cross-validation~\cite{the-k-in-k-fold-cross-validation}, and stratified sampling~\cite{stratified-sampling-meets-machine-learning}. Due to the large amounts of potential I/O and computation required to prepare the data effectively, parallelization techniques are needed to ensure proper system utilization. ML applications typically use the parallelization tools provided by their primary ML framework (e.g., PyTorch~\cite{pytorch-docs} and TensorFlow~\cite{tensorflow-docs}) to parallelize data processing. Once the data has been prepared, it can be fed to the model for training, validation, testing or inference.

During the \textbf{training} phase, data from the training dataset is read into memory and fed to the model. The order in which the data is fed to the model is largely impacted by the optimization algorithm of choice. Gradient descent algorithms are a popular set of optimization algorithms used to train ML models. Due to stochastic gradient descent (SGD) being the most popular gradient descent algorithm for large datasets~\cite{recent-trends-in-stochastic-gradient-descent-for-machine-learning-and-big-data}, the algorithms we will present are batch gradient descent (BGD)~\cite{theoretical-analysis-of-batch-and-on-line-training-for-gradient-descent-learning-in-neural-networks}, SGD~\cite{stachastic-gradient-descent-tricks} and mini-batch gradient descent (MBGD)~\cite{optimization-methods-for-large-scale-machine-learning}. SGD is a specific case of MBGD, and in the future, we will refer to both as SGD unless a distinction is necessary. Deciding which algorithm to use depends on the dataset size, the computational resources, and the underlying model's properties. It is important to note that gradient descent algorithms vary in their mathematical properties~\cite{on-the-generalization-benefit-of-noise-in-stachastic-gradient-descent}. At every epoch, each of the listed algorithms feeds the entirety of the dataset to the model. However, the frequency at which the model is updated, and the order in which the dataset is fed to the model differ.

BGD is the most straightforward gradient descent algorithm. This technique is computationally expensive and is often used only when the entire dataset can fit into memory. The model is trained against the whole dataset before it can perform one update~\cite{gradient-descient-algorithms-analysis}. If the model cannot cache the entire dataset in a fast storage tier, it necessitates utilizing underlying storage tiers, leading to increased I/O operations. MBGD is an algorithm where random subsets or batches are chosen and fed to the model. The batch size varies depending on the specific model being used. However, if the batch size is small, the model becomes susceptible to noise updates (updates that introduce randomness or variability in the parameter updates), while large batch sizes, in contrast, may lead to over-fitting. The SGD algorithm involves updating the model after feeding a single data point. SGD is a specific case of MBGD where the batch size is $1$. The random nature of the data selection when using SGD makes it difficult to cache the same mini-batches or data points from previous epochs for future ones. Each epoch involves a new random sampling of the data, introducing variability in the training process. This commonly leads to an I/O pattern of large amounts of small I/O performed on many random files. When training models on an HPC platform, datasets are located on a shared Parallel File System (PFS), designed for reading large batched files rather than a large number of small files. This poses a significant challenge for PFSs~\cite{i-o-characterization-and-performance-evaluation-of-beegfs-for-deep-learning}.


Finally, after a certain number of epochs, it is common practice to assess the model's performance while tuning hyperparameters using a separate validation dataset~\cite{pattern-recognition-and-neural-networks}. The same validation dataset to often used after each epoch to evaluate the model. As the training and validation phases are time-intensive stages within the ML lifecycle, distribution strategies have been developed to parallelize these tasks. The distribution strategies optimize the utilization of computational resources, thereby expediting the training process and enhancing efficiency in model development. Choosing the appropriate training and validation distribution strategy depends on factors such as model size, data set size, and available computational resources. This decision is pivotal in guaranteeing optimal training performance. It should be noted that after the model has converged, there is sometimes an additional phase called the \textbf{testing} phase which evaluates the performance of the model on unseen data using the testing dataset.

\subsection{Distributing the Training and Validation Phases}
\label{subsection:optimizing-training-evaluation-phases}

\begin{figure}[!t]
    \includegraphics[width=15cm]{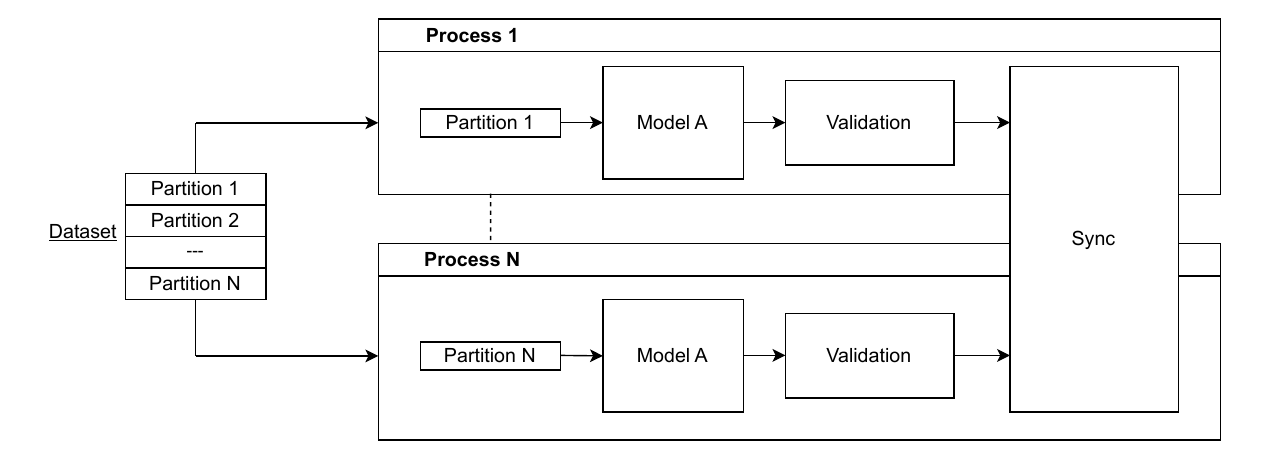}
    \caption{Data parallelism where the dataset is broken into $N$ partitions. Each partition is then fed to the model in $N$ processes. After validation, all processes must synchronize with each other.}
    \label{fig:data-parallelism-diagram}
\end{figure}

\begin{figure}[!t]
    \includegraphics[width=15cm]{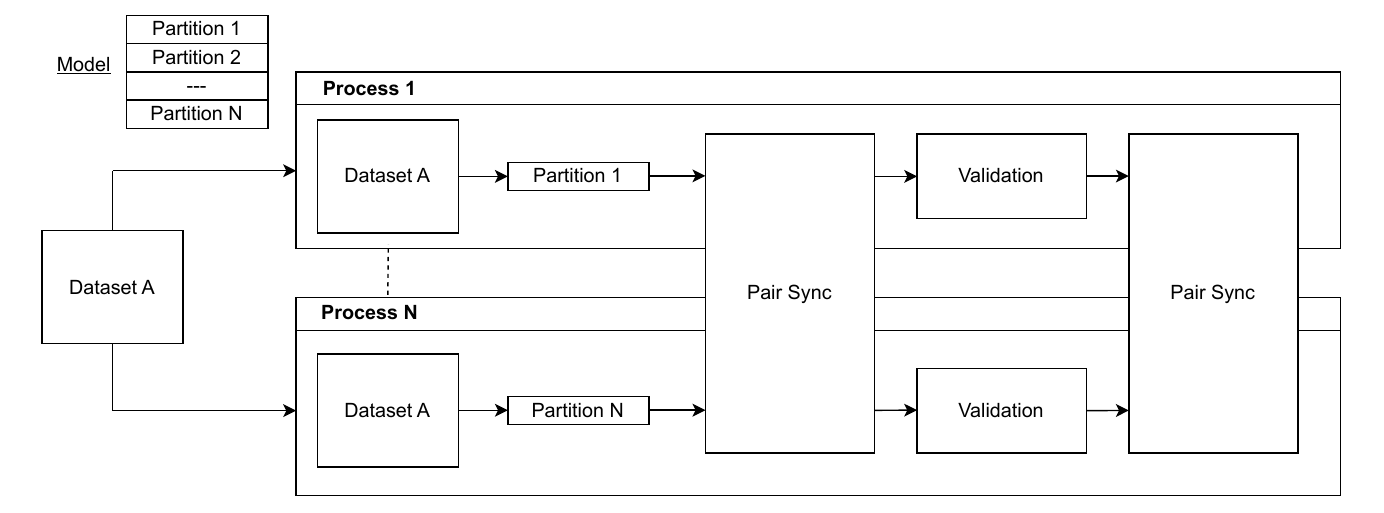}
    \caption{Model parallelism where the model is broken into $N$ partitions. Pairs of processes synchronize after both training and validation. Each model partition is trained against the entire dataset at each epoch.}
    \label{fig:model-parallelism-diagram}
\end{figure}

Multiple strategies exist to distribute the training and validation phases to take advantage of the parallelism granted by HPC systems. Two common techniques used are \textbf{model parallelism} and \textbf{data parallelism}, shown in Fig.~\ref{fig:data-parallelism-diagram} and Fig.~\ref{fig:model-parallelism-diagram}, respectively. When using model parallelism, a single model is distributed across $N$ workers, and each worker trains their partition of the model~\cite{towards-accelerating-model-parallelism-in-distributed-deep-learning-systems}. It is often used when the model is too large to fit into the memory of a single node. The strategy has two pair synchronization points: one before the validation and another after adjusting each process's model partition. Synchronization occurs among pairs of workers and is completed across multiple steps. This allows pairs of workers who have been updated to continue training without having to synchronize with all other workers. When using data parallelism, $N$ workers are employed, and the model is replicated $N$ times across each worker which each train against a different dataset partition. After each iteration, all workers must be synchronized, and the model weights are adjusted according to the results of the backpropagation. This is the parallelization strategy used in Section~\ref{subsection:training_io_access_patterns} to analyze the I/O patterns observed during training. In most cases, model parallelism's synchronization and communication overhead far exceeds data parallelism's~\cite{an-efficient-algorithm-for-data-parallelism-based-on-stochastic-optimization}. However, both model parallelism and data parallelism involve inter-device (i.e., TPU, GPU, or CPU) communication to exchange information about model parameters and gradients. This communication overhead can impact training performance. Strategies such as gradient accumulation, where model updates are aggregated over several batches before communication or gradient compression~\cite{adacomp-adaptive-residual-gradient-compression-for-data-parallel-distributed-training} can help reduce the overhead of inter-device communication. 

More recent training distribution strategies include hybrid parallelism~\cite{extrememly-large-minibatch-sgd}, and pipeline parallelism~\cite{pipline-parallelism}. Hybrid parallelism combines data and model parallelism, where the data and the model are partitioned across multiple devices. In contrast, pipeline parallelism places different layers of the model on separate devices, allowing for efficient overlapping of computation and communication. Federated learning, introduced by Google in 2016, is a training distribution strategy where the model is partitioned into multiple devices which train their model partition against private data~\cite{federated-learning-opportunities-and-challenges, communication-efficient-learning-of-deep-networks-from-decentralized-data}. After some specified amount of training, devices send their model gradients to a central server, which aggregates the model gradients and redistributes the results back to the devices. This is beneficial for use cases in which the data being used to train the model is sensitive, such as in healthcare or finance. The training data is no longer aggregated into one central location resulting in an increase in privacy and security. The sending and receiving of the model gradients with a central server can result in significant network communication overhead. Various optimizations can reduce communication overhead, such as intelligently selecting the nodes or devices that participate in the training, semi-asynchronous training methods, and clustered federated learning: a strategy in which groups of nodes are prioritized over others during training based on various factors such as latency or physical distance~\cite{async-federated-learning-on-hetergenous-devices}.

\subsection{Model Checkpoints}
\label{subsection:model_checkpoints}


While large amounts of small reads dominate the I/O operations performed when training ML models, a consistent pattern of writes also occurs~\cite{analyzing-the-io-patterns-of-deep-learning-applications}. This is due to the need to checkpoint the model as it is being trained. This can significantly impact the training performance, especially if the checkpoints are done frequently. Checkpoints are important as they allow long-running training jobs to pick up where they left off in the event of node failures. In addition, it allows for the analysis of intermediate model features in relation to the training data. Many training jobs begin from an existing model checkpoint from which fine-tuning for their specific task can be accomplished. Starting from a pretrained model can reduce the time and data needed for training. When checkpointing a model, the write size depends largely on the number of parameters the ML model requires~\cite{optimizing-asynchronous-multi-level-checkpoint-restart-configurations-with-machine-learning}. Because of the increasing complexity of ML models~\cite{machine-learning-model-sizes-and-the-parameter-gap}, checkpoint sizes can be expected to grow proportionally. Most ML frameworks use $32$ bit precision to store model parameters~\cite{efficient-processing-of-deep-neural-networks-a-tutorial-and-survey}, the default parameter size on both PyTorch and TensorFlow. Large language models (LLM)s such as GPT-3 and BERT large have $175$ billion~\cite{language-models-are-few-shot-learners} and $345$ million~\cite{bert-pre-training-of-deep-bidirectional-transformers-for-language-understanding} parameters respectively. The time needed to checkpoint the model depends not only on the number of parameters but also on the file format used~\cite{a-study-of-checkpointing-in-large-scale-training-of-deep-neural-networks}, making it an important consideration for training performance. In addition, when using frameworks such as TensorFlow, checkpointing is done independently by each worker, or a single worker is assigned the job. This can lead to the stragglers problem due to a single process being responsible for the extra work. When using a single worker to checkpoint the model, the time required does not shrink as more nodes are added to the training. PyTorch only recently (October $4$, $2023$ in version $2.1$)  added support for distributed checkpointing, allowing for the saving and loading of models to be accomplished in parallel by multiple ranks. Efforts to optimize model checkpointing include model compression~\cite{on-efficient-construction-of-checkpoints} and asynchronous checkpointing. A description of the popular ML frameworks and their corresponding checkpoint APIs can be found in Section~\ref{subsection:ml_frameworks}.

\subsection{BERT Real-world End-to-end ML Workflow} 

In this section, we present the well-known real-world end-to-end ML workflow of BERT. We discuss BERT in place of GPT because GPT is a closed source, meaning the details about the exact I/O found throughout its lifecycle are not fully disclosed. The authors of BERT mention that it has the most comparable existing pre-training method to OpenAI GPT. The information presented in this section is based on both the original BERT paper and the updated open-source GitHub repository, which specifies important data management attributes such as file formats and pre-training times. A more detailed analysis of the BERT training I/O patterns can be found in Section~\ref{subsection:training_io_access_patterns} where we simulate the BERT workload using the DLIO~\cite{dlio-a-data-centric-benchmark-for-scientific-deep-learning-applications} benchmark.

\subsubsection*{Data Generation} \ The data used to pre-train the model is the BooksCorpus dataset~\cite{bookscorpus-dataset}, which consists of $11,000$ books with a total of $800$ million words, and the English Wikipedia consisting of $2,500$ million words.

\subsubsection*{Offline Data Preparation} \ BERT is an unsupervised LLM, meaning it is trained on a large quantity of publicly available plain text data. The Wikipedia data is stripped of all texts other than text passages, ignoring lists, tables, and headers. The text from both datasets is processed into a set of input token sequences, which can then be fed to the model during pre-training. The input datasets are transformed into a single large plain text file, referred to as the corpus, with each sentence being separated by a new line. It is noted that the newline-delimited text sequences contained within the corpus should be actual sentences to enable ``next sentence prediction.'' Token segmentation is then used to divide the text into individual tokens using a subword tokenizer named WordPiece~\cite{wordpiece}. A distinguishing factor of BERT is that it uses bidirectional contextual representations, i.e., a given word's representation is based on the words found to its left and right. This is accomplished by masking out a random $15$\% of the tokens within each token sequence. The masked-out tokens are then used as the prediction target for the sequence. The token sequence representation used is of the form \texttt{<Question, Answer>}, allowing for both a single sentence and a pair of sentences. This is done by placing special tokens such as \texttt{[SEP]} representing a sentence delimiter and \texttt{[CLS]} representing a token sequence starting point. The \texttt{[SEP]} token delimiter is used to separate different segments of text (e.g., question and answer pairs). The output of the data generation is a set of TFRecord files containing the token sequences.

\subsubsection*{Pre-training} \ The Tensorflow framework is used to train the model during pre-training. This step in the workflow takes around $4$-$6$ days when using $4$ to $16$ Cloud TPUs. However, it is a one-step procedure from which specific versions of the models can be based on during fine-tuning. Notable hyperparameters used to pre-train BERT in the paper are a batch size of $256$ sequences where each sequence is $512$ tokens, resulting in $128,000$ tokens per batch. During pre-training, random sentences are chosen in pairs \texttt{A} and \texttt{B} in which $50$\% of the time sentence \texttt{B} is the sentence following \texttt{A}. This is done to enable ``next sentence prediction.''

\subsubsection*{Fine-tuning} \ The fine-tuning results presented are relatively inexpensive and can be replicated in at most one hour on a single TPU or a few hours in a single GPU. All results presented in the paper were fine-tuned on a single TPU due to memory constraints. To begin, the model is loaded from a previously generated checkpoint generated from the pre-training phase. The checkpoint consists of three files containing the model weights. The hyperparameters used for fine-tuning are similar to pre-training other than batch size, learning rate, and number of epochs. The paper recommends training for $2$, $3$, or $4$ epochs using a batch size of $16$ or $32$ and a learning rate of $5$d-$5$, $3$e-$5$, or $2$e-$5$. Training with long sequence lengths in constrained memory environments can result in out-of-memory errors because the attention method's memory requirement increases by the square of the sequence length.

\subsubsection*{Evaluation} \ After a series of pre-training and training steps, an optional validation or testing phase can be enabled or disabled through the \texttt{-do\_eval} parameter. This phase is used to track model performance on validation or testing datasets, fine-tune hyperparameters, or select the best model to checkpoint.

\subsubsection*{Inference} \ When using BERT for inference, data preprocessing steps similar to those described above must be done. During either offline or online inference, text input must be tokenized and processed into token sequences. The distinguishing factor of online inference is that the token sequences are fed to the model, and their associated predictions must be processed in real-time. Due to the wide range of possible deployment scenarios, categorizing the data management and I/O that occur in this phase is challenging. However, notable factors distinguishing this phase are the need to store or stream the prediction generated from the model and that there is no longer a need to checkpoint the model.

\begin{tcolorbox}[colback=white!5!white,colframe=black!75!black,title=Summary $\#2$]
ML applications use a variety of file formats and dataset modalities. The dataset modality impacts what file format is used, affecting how samples are stored and accessed and what features are available. These choices often impact the performance and ease of use in accessing data for ML analysis. Furthermore, during the training phase, SGD is the most popular algorithm in which random batches of samples are fed to the model. The random nature of SGD and the parallelism resulting from training distribution methods (i.e., model and data parallelism) cause I/O operations that are challenging for PFSs. Finally, with the increasing complexity of models, it is crucial to ensure efficient checkpointing methods to achieve fast training speeds.
\end{tcolorbox}

\section{I/O Benchmarks and Profiling}
\label{sec:ml-io-profiling-and-benchmarks}


Many ML benchmarks have been proposed, such as Fathom~\cite{fathom-reference-workloads-for-modern-deep-learning-methods} and BenchNN~\cite{benchnn-on-the-broad-potential-application-scope-of-hardware-neural-network-accelerators}. However, these tools focus on identifying the computational requirements of ML workloads and not the I/O requirements~\cite{dlio-a-data-centric-benchmark-for-scientific-deep-learning-applications}. In addition, they target cloud platforms. 
This leaves a need for a proper benchmark focusing on I/O that simulates ML workloads.

One attempt to address this gap is DLIO~\cite{dlio-a-data-centric-benchmark-for-scientific-deep-learning-applications}, a benchmark designed to simulate I/O access patterns commonly found in Deep Learning (DL) workloads. DLIO, which has been released as MLPerf Storage Benchmarks~\cite{ml-perf-inference-benchmark}, allows for extensive configuration, including the selection of interfaces (HDF5, TFRecord, CSV, NPZ), file access patterns (one file per process versus shared file per process), data access patterns, I/O types, and transfer buffer sizes. Scientists and researchers can take advantage of DLIO by configuring the benchmark to mimic the expected I/O patterns of their DL workloads and using the results to configure their applications effectively. DLIO also supports DALI~\cite{nvidia-dali}, which is capable of sending batches to the GPU for preprocessing and model training, primarily for image or video datasets. Current DLIO limitations include the inability to store multiple samples per file in certain file formats, the inability to store data/label pairs in separate files, and the assumption that the samples will always be 2D images. The inability to store multiple samples per file poses a challenge when comparing workloads across ML frameworks. For example, the BERT workload uses the TFRecord format, but currently, there is no PyTorch data loader capable of loading TFRecords. Transitioning to an alternative file format requires configuration modifications, resulting in a workload that diverges from the original and may not faithfully represent it. Thus, while DLIO effectively captures a broad spectrum of DL applications, comparing different frameworks on the same workload while maintaining accuracy proves challenging. 

Cheng et al.~\cite{storage-benchmarking-with-deep-learning-workloads} benchmark two object storage systems (MinIO and Ceph) and three key-value databases (MongoDB, Redis, and Cassandra). Their study attempts to find optimal configurations for parameters such as storage location, storage disaggregation granularity, access pattern, and data format. Notable takeaways are that the object storage systems tested were more sensitive to storage location than key-value databases. In addition, the object storage systems tested were impacted mainly by the storage disaggregation granularity as only one object could be queried at a time. 

Many dataset modalities, such as audio, video, and images, have a surplus of public datasets to choose from. In contrast, high-quality, large-scale graph datasets are hard to find; in addition, most commonly used graph datasets are too small to emulate the datasets found in the real world.  Hu et al.~\cite{open-graph-benchmark} have created Open Graph Benchmark (OGB), which aims to enhance graph ML research by providing robust and scalable graph datasets. It is capable of producing graph datasets of three different size categories: small, medium, and large, which have $100,000$ nodes, $1,000,000$ nodes, and $1,000,000,000$ nodes, respectively. OGB enables researchers to analyze large-scale graph datasets, facilitating rapid growth in distributed graph machine learning (GML).

iBench~\cite{ibench} is a possible solution for distributed inference simulation and benchmarking and aims to provide several key metrics, including but limited to throughput, latency, ingest rate/bandwidth, pre-processing time, and GPU efficiency. Compared to MLPerf, they note two primary advantages: (1) the ability to measure distributed inference performance and (2) a more realistic performance measure. The simulation places a stream load balancer in front of the HPC inference servers. Client requests are then simulated and scheduled to run via the load balancer. They aim to accurately assess performance by collecting significant factors such as inference time, network response time, payload preprocessing and packing time, and investment time. Due to the limited number of available streaming HPC inference benchmarks, iBench is a promising tool for evaluating network I/O. Current limitations of iBench include the inability to represent many ML models and deployable HPC edge devices. Brewer et al.~\cite{ibench-evaluation} evaluated iBench, noting several key takeaways, including the impact on performance when using different client batch sizes of $16$, $32$, or $64$, and processing requests asynchronously, the importance of topology-aware GPU-TPU and CPU-GPU communication, and the significance of CPU-GPU in inference compared to training. The evaluation concludes that linear scaling can be achieved if multiple client requests are processed asynchronously and the amount of data sent does not saturate the network.

While benchmarking tools allow ML developers to emulate various ML workload I/O patterns, profiling tools that enable developers to analyze the HPC I/O stack in its entirety are crucial for pinpointing potential bottlenecks. TensorFlow offers the TensorFlowProfiler, released in version $2.2.0$, capable of profiling host-side and GPU tracing. However, it only provides information at the TensorFlow level, not capturing lower-level I/O operations such as POSIX or STDIO. Chien et al.~\cite{tf-darshan-understanding-fine-grained-i-o-performance-in-machine-learning-workloads} created tf-Darshan, a fine-grained I/O profiling tool. It relies on Darshan~\cite{modular-hpc-i-o-characterization-with-darshan}, a highly popular HPC I/O profiler and tracer. Tf-darshan augments Darshan, enabling the runtime extraction and analysis of Darshan's internal data structures. Runtime extraction is accomplished through an augmentation of the Darshan library. It adds extraction functions that return its internal buffers that store information, such as record counters for I/O operations, to individual files. This gives ML applications the ability to perform runtime analysis, enabling I/O optimization techniques such as specialized online I/O schedulers to perform runtime optimizations. Tf-darshan also offers a visualization tool that allows users to understand the I/O performance of their ML workloads intuitively. Current limitations include exclusive TensorFlow integration and a moderate overhead of $10\%$ - $20\%$.

\begin{tcolorbox}[colback=white!5!white,colframe=black!75!black,title=Summary $\#3$]
Benchmarks play a crucial role in accelerating research within ML I/O. They enable the identification of I/O bottlenecks in ML workloads, thereby enhancing efficiency and productivity. Due to the increase of heterogeneity found in HPC architectures, profilers capable of capturing data movement between a wide range of devices and data sources (e.g., memory, disk, archive, cloud, or database) would be beneficial. Furthermore, existing benchmarks fail to represent the I/O patterns found during all stages of the ML lifecycle, highlighting the need for benchmarks capable of emulating I/O at various stages and data sources.
\end{tcolorbox}

\section{Analysis of I/O Access Patterns}
\label{sec:ml-io-access-patterns}

In this section, we aim to present the I/O access patterns commonly found during distributed offline data preparation, training, and inference. We searched for benchmarks and found DLIO~\cite{dlio-a-data-centric-benchmark-for-scientific-deep-learning-applications} to be the most promising to simulate the training I/O access patterns of DL applications in HPC systems.

\subsection{Offline Data Preparation Access Patterns}

During the offline data preparation phase, I/O access patterns differ depending on the specific operations needed to prepare the dataset for training or inference. Common operations such as data transformations, normalization, or feature scaling can often be distributed across compute nodes. Due to the high complexity that arises in the programming of distributed systems applications, big data frameworks such as Apache Spark~\cite{spark-docs}, Apache Hadoop~\cite{apache-hadoop}, or Dask~\cite{dask-docs} are often utilized to easily distribute data preparation tasks across nodes. Several popular file formats, including .json, .tfrecord, and .npz, do not natively support random I/O access~\cite{accelerating-deep-learning-training-on-hpc-systems}. Instead, data must be read sequentially, enforcing sequential I/O while eliminating parallelism. These file formats are often ideal for streaming scenarios while hindering dataset preparation operations such as selections and filtering. To enable parallel processing for file formats that do not offer random I/O access, the dataset is often sharded, that is, divided into smaller files. Sharding enables distributed processing among compute nodes on each of the ''shards`` of the dataset. This often results in each process sequentially reading a shard of the dataset, increasing I/O throughput. Although sharding is optional and may require additional processing time, it is often worth the trade-off in reduced I/O times~\cite{high-performance-io-for-large-scale-deep-learning}. In general, sharding and the use of big data frameworks provide essential tools to enable ideal I/O access patterns and effective use of the HPC system.

\subsection{Training I/O Access Patterns}
\label{subsection:training_io_access_patterns}

\begin{figure}
    \centering
    \subfigure[One Sample per File]{\includegraphics[width=0.4\textwidth]{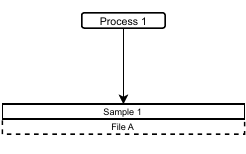}} 
    \subfigure[Multiple Samples per File]
    {\includegraphics[width=0.4\textwidth]{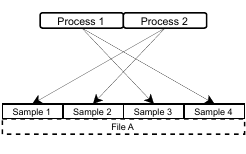}} 
    \caption{Process level accesses for common organizations of samples within files.}
    \label{fig:samples-per-file}
\end{figure}

The I/O access patterns found during the training phase are influenced by the gradient descent algorithm used and the training distribution strategy. Here, we describe the I/O access patterns found when using SGD, because SGD is the most popular gradient descent algorithm when training with large datasets. In addition, the presented workloads use data parallelism, where the model is replicated in each worker and the data is partitioned and distributed among them~\cite{analyzing-the-distributed-training-of-deep-learning-models-via-data-locality}. The organization of the samples within the files, as seen in Fig.~\ref{fig:samples-per-file}, has an impact on the I/O operations performed by each worker. Two common organizational strategies are \textit{one sample per file}, often used with file formats such as JPEG or WAV, and \textit{multiple samples per file}, commonly used with file formats such as TFRecord or HDF5. The choice of sample organization plays a large role in the observed I/O operations~\cite{a-case-study-of-data-management-challenges-presented-in-large-scale-machine-learning-workflows}. A significant amount of additional metadata management overhead can occur when using many files \cite{high-performance-io-for-large-scale-deep-learning}. The Luster PFS used in the workloads presented below has multiple metadata servers (MDS)s which are responsible for providing metadata services such as filename lookup, directory information, file layouts, and access permission. When using \textit{one sample per file} with a dataset with many samples, MDSs can be overwhelmed by concurrent file requests, causing slower training speeds. We show both sample organizations with the Unet3D workload (\textit{one sample per file}) and the BERT workload (\textit{multiple samples per file}).

We obtained the results below by running DLIO and captured the I/O traces using Darshan~\cite{snyder:2016:darshan}. All benchmarks were conducted with Perlmutter's compute nodes. Perlmutter is a supercomputer located at the National Energy Research Scientific Computing Center of the United States Department of Energy. Each of the compute nodes has $64$ physical cores and $128$ virtual cores, with a total memory capacity of $255$ GiB. The PFS is an all-flash Luster file system, containing a total bandwidth of more than $5$ TB/sec and supporting $4$ million IOPS ($4$ KiB random). Perlmutter has $16$ MDSs and 298 object storage targets (OST)s, and all workloads presented use a $1$ MiB stripe size. OSTs are storage devices in which the user's file data is stored. For visualization, we used DXT Explorer~\cite{dxt-explorer}, which is an interactive web-based tool capable of producing graphs from Darshan DXT logs.


\begin{figure}[htbp]
\includegraphics[width=\textwidth]{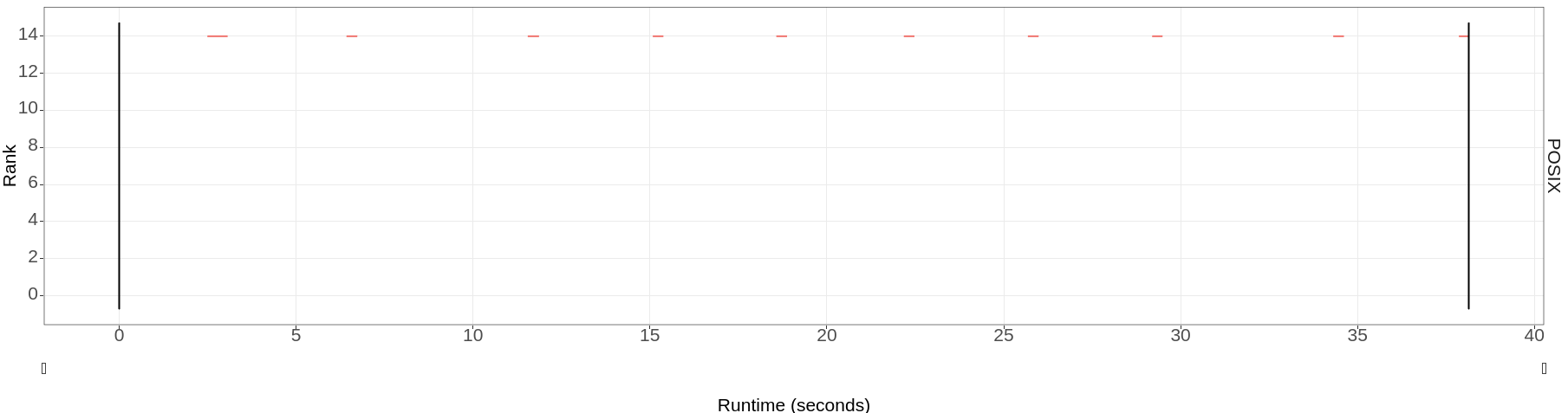}
\caption{DLIO TensorFlow reads during the Unet3D workload from the view of an individual file.}
\label{fig:unet3d_reads_size_time}
\centering
\end{figure}

In Fig.~\ref{fig:unet3d_reads_size_time}, we used DLIO to simulate the training of Unet3D, which is used for 3D medical image segmentation tasks. The workload consisted of $4$ compute nodes, each running $4$ Python processes. Python processes were given access to $4$ physical CPU cores, and the TensorFlow data loader was configured with $4$ I/O threads and a batch size of $4$ samples. The dataset for Unet3D consists of one sample per NPZ file where each sample is approximately $146$ MiB. There were a total of $168$ samples. The model was trained for $10$ epochs. Due to the one-to-one correspondence between the number of samples per file, the entire file is requested by the same process at each epoch rather than subsets of the file being requested. In addition, the file was accessed by the same rank ($14$ in this case) at each epoch.


\begin{figure}[!t]
\includegraphics[width=\textwidth]{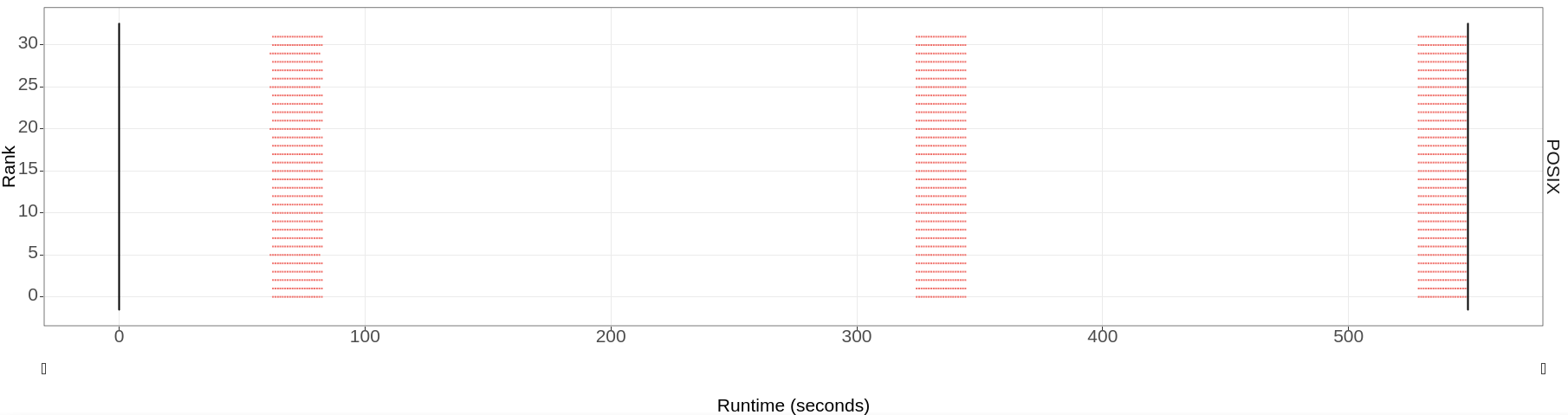}
\caption{DLIO TensorFlow reads during the Bert workload from the view of an individual file.}
\label{fig:bert_reads_ranks_time}
\centering
\end{figure}

In Fig.~\ref{fig:bert_reads_ranks_time}, we used DLIO to simulate the training of BERT, which is a large language model (LLM). The computational setup consisted of $8$ compute nodes, each running $4$ Python processes. Python processes were given access to $2$ physical CPU cores, and the TensorFlow data loader was configured with $1$ I/O thread. The dataset size was reduced to increase training speeds; however, we do not expect this to significantly change the overall data access patterns. We reduced the number of files from $500$ to $10$, and the number of samples per TFRecord file from $313,532$ to $31,353$. Each sample was $2,500$ bytes. Therefore, there were $10 \times 31,353 = 313,530$ samples in total. The workload was run twice, once with the original batch size of $48$ and a second time with a batch size of $96$. The model was trained for three epochs. 

\begin{figure}[!b]
\includegraphics[width=\textwidth]{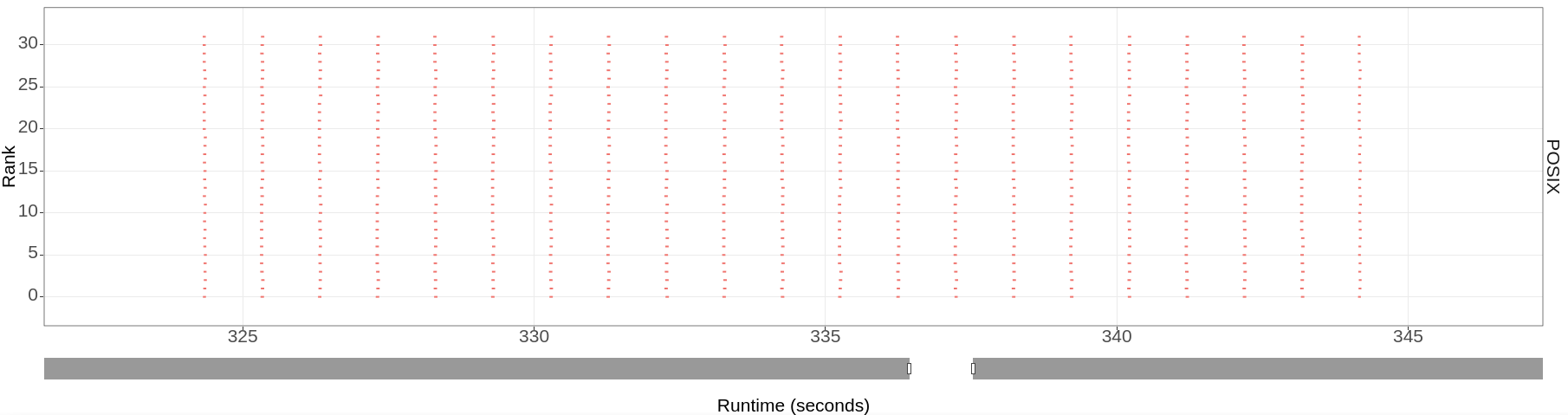}
\caption{Reads during the BERT workload zoomed in on the second epoch of Fig.~\ref{fig:bert_reads_ranks_time}.}
\label{fig:bert_reads_ranks_time_single_epoch_zoom}
\centering
\end{figure}

\begin{figure}[htbp]
\includegraphics[width=\textwidth]{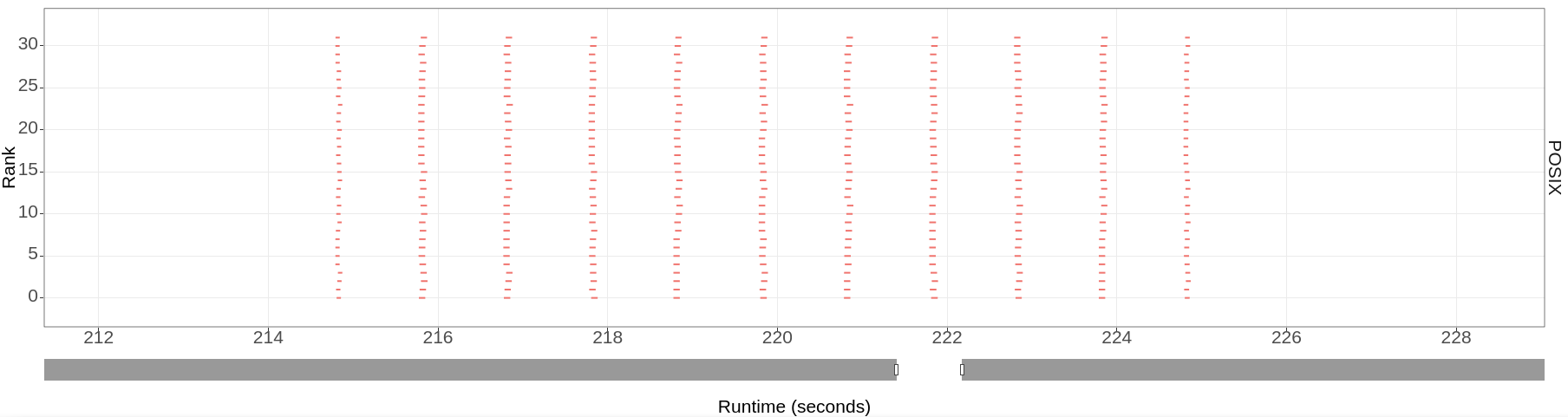}
\caption{Reads during the BERT workload zoomed in on the second epoch where batch sized was doubled to be $96$ samples.}
\label{fig:bert_reads_ranks_time_single_epoch_zoom_increased_batch_size}
\centering
\end{figure}

Fig.~\ref{fig:bert_reads_ranks_time_single_epoch_zoom} is a zoomed-in view of the second epoch of Fig.~\ref{fig:bert_reads_ranks_time}. There are $21$ distinct columns of I/O reads that occur at each epoch from every process. This can be expected because the result of $\frac{total\_samples}{number\_of\_processes \cdot samples\_per\_batch \cdot total\_files}$ would give us the number of batches that need to be read by every process each epoch with multiple samples per file. Taking into account the $313,530$ samples, the $48$ samples per batch and the $32$ MPI processes, there are $\frac{313,530}{32 \cdot 48 \cdot 10} = \left\lceil 20.412... \right\rceil = 21$ batches of samples to read per epoch (rounding up to the nearest integer as the remaining samples need to be read). To ensure this result, we ran the BERT workload again but doubled the batch size to $96$ samples. Fig.~\ref{fig:bert_reads_ranks_time_single_epoch_zoom_increased_batch_size} is a zoomed-in perspective on the second epoch and can be seen to have $11$ distinct columns of I/O reads which agrees with $\frac{313,530}{32 \cdot 96 \cdot 10} = \left\lceil 10.206... \right\rceil = 11$ batches of samples to read per epoch. The non-overlapping nature of the columns is expected, because all processes synchronize after every batch is read. 

Since no caching is used for the workloads presented, every read request must reach the underlying PFS for every sample needed. The absence of caching poses a potential I/O bottleneck, particularly when dealing with large sample sizes or a large number of small samples. The dividing of the datasets among the ranks and the subsequent I/O reads for batches of samples aligns with the observations of other studies documenting ML training I/O access patterns~\cite{why-globally-re-shuffle-revisiting-data-shuffling-in-large-scale-deep-learning, dlio-a-data-centric-benchmark-for-scientific-deep-learning-applications, storage-benchmarking-with-deep-learning-workloads}. The interaction between data organization strategies and the underlying PFS significantly influences the performance, with potential bottlenecks emerging from high volumes of small reads or extensive metadata operations. Understanding and optimizing these patterns is essential for enhancing the efficiency of training ML models on HPC systems.

\subsubsection{Checkpoints}
\ As mentioned in Section~\ref{subsection:model_checkpoints}, it is common for a consistent pattern of writes to occur during the training loop to checkpoint the model. Both the Unet3D and BERT workloads use the TensorFlow \texttt{\small checkpoint} function described in Section~\ref{subsection:tensorflow_api}, where each checkpoint consists of three files: an index file, a metadata file, and a data file. In both workloads, within the checkpoint directory, a file named ``checkpoint'' has the most recent checkpoint version. For the Unet3D workload, checkpointing was set to occur after epoch $5$, then every $2$ epochs thereafter. This resulted in a total of $3$ checkpoints after epochs $5$, $7$, and $9$. For BERT, checkpointing was set to occur after every $250$ steps. Here, we describe the number of checkpoints when using a batch size of $48$. Since the total number of samples was $313,530$, the model was trained for $3$ epochs, and checkpoints occur every $250$ steps, the total number of checkpoints was $\left\lfloor \frac{3 * 313,530}{48* 250} \right\rfloor = 78$. As observed, frequent checkpointing can result in a moderate number of files and, consequently, increased write operations. While checkpointing is essential for fault tolerance and recovery, it also introduces additional I/O overhead that may impact training efficiency. ML developers must strike a balance between ensuring node failure resilience through checkpointing and maintaining reasonable I/O performance to avoid training bottlenecks.

\subsection{Inference I/O Access Patterns}

In this section, we aim to categorize the most common I/O access patterns found during the inference phase within HPC systems. MLPerf~\cite{ml-perf} developed by MLCommons is a well-known group aiming to provide ML training and inference evaluations. They categorize ML inference into four categories: single-stream, multi-stream, server, and offline~\cite{ml-perf-inference-benchmark}. However, these inference categories were not specific to HPC environments, and two of the categories (single-stream and server) were not found in the cited literature or in our own experience. We have therefore excluded these categories from this discussion and have found the two categories described below to be representative of most inference configurations found in HPC systems.

\begin{enumerate}
    \item Single-stream: multiple client devices stream single sample queries to the HPC system in parallel.
    \item Offline: data is immediately available, and latency and bandwidth are unconstrained, e.g., data is stored in a PFS.
\end{enumerate}

A key difference between streaming and offline inference is the system driving the inference process. In the offline scenario, the HPC system requests data from the storage system, whereas, in streaming scenarios, the HPC system is driven by end users or clients. Latency and bandwidth become more constrained in streaming scenarios, which requires effective load balancing to manage unpredictable query request times to effectively utilize compute nodes~\cite{ibench, ibench-evaluation}. A significant amount of research has been done in this area when ML models are deployed on edge computing devices where constraints such as limited memory, high latency, and limited bandwidth are common~\cite{resource-allocation-with-edge-computing-in-iot-networks-via-machine-learning, edge-ai-on-demon-accelerating-deep-neural-network-inference-via-edge-computing, pipedge-pipline-parallelism-for-large-scale-model-inference-on-heterogeneous-edge-devices}. Data reduction techniques such as compression~\cite{deep-compressive-offloading-speeding-up-neural-network-inference-by-trading-edge-computation-for-network-latency} and intelligent grouping of edge computing resources~\cite{resource-allocation-with-edge-computing-in-iot-networks-via-machine-learning} help mitigate network I/O constraints. Although some of the optimizations mentioned above can be used in offline inference, lower latency and higher bandwidth are often found within HPC systems due to high-speed interconnects.

When inference is compared with training, several distinctions emerge. During inference, data is often not sampled randomly, which can result in an I/O access pattern of large contiguous reads in scenarios such as offline inference. High throughput can be expected during offline inference due to PFSs excelling in large contiguous readings. In a streaming scenario, contiguous reads may not be possible if the data is not readily available. Additionally, checkpointing is less common in inference because the model is no longer updated. It should be noted that there are some scenarios, such as online learning and adaptive inference, that may still involve model updates and checkpointing. In addition to checkpointing, storing the prediction results is another potential source of I/O writes. In a streaming scenario, the prediction results may need to be sent back to the querying client, causing more contention on the network bandwidth.

\begin{tcolorbox}[colback=white!5!white,colframe=black!75!black,title=Summary $\#4$]
Due to the prevalence of SGD, random batches of samples are read into memory at each iteration during model training. Small random I/O reads can be a bottleneck for PFSs, which motivates the need for I/O optimization techniques such as prefetching and caching. Inference involves a diverse range of scenarios. When streaming over the wide-area network, latency and bandwidth constraints are of greater concern. In an offline scenario, large contiguous reads are possible, which perform favorably on traditional HPC storage systems.
\end{tcolorbox}

\section{I/O Optimization Techniques}
\label{sec:ml-io-optimization-techniques}

This section discusses commonly used ML frameworks and the I/O optimization options developers can utilize to distribute their dataset preparation and training phases. Based on the current research, we also discuss potential I/O optimization techniques ML workloads can use to increase training speeds.

\begin{table}[h]
\centering
\begin{tabular}{lccc}
\toprule
\multirow{2}{*}{\textbf{I/O Feature}} & \multicolumn{2}{c}{\textbf{ML Frameworks}} \\
\cmidrule(lr){2-3}
& PyTorch & TensorFlow \\
\midrule
Dataset Streaming & \YES & \YES \\
Sample Prefetching & \YES & \YES \\
Sample Caching & \NO & \YES \\
Distributed Sample Caching & \NO & \YES \\
Asynchronous Checkpointing & \NO  & \NO \\
Distributed Checkpointing & \YES & \NO \\
\bottomrule
\end{tabular}
\caption{I/O features from popular ML frameworks.}
\label{tab:features-from-popular-ml-frameworks}
\end{table}

\subsection{Common I/O Optimization Techniques Used by Current ML Frameworks}
\label{subsection:ml_frameworks}


Here, we explore the data loading and preprocessing capabilities offered by different ML frameworks. We describe the various classes and methods in state-of-the-art frameworks, including PyTorch, TensorFlow, and Scikit-learn with Dask-ML. Additionally, we touch upon the usage of tools, including Horovod for distributed deep learning across various platforms and direct NVIDIA GPU loading for direct loading from storage and network devices to the GPU. The I/O features supported by each ML framework can be seen in Table \ref{tab:features-from-popular-ml-frameworks}. It should be noted that ML frameworks often aim to be generally applicable and extendable, meaning that it is often possible to extend one framework to have a feature of another. For example, sample caching can be easily integrated into PyTorch even though it does not offer this option through its API.  Table~\ref{tab:features-from-popular-ml-frameworks} only shows a checkmark for a feature if it is easily enabled through the API without extending the underlying framework significantly.

\subsubsection{PyTorch} \ PyTorch~\cite{pytorch-docs} offers the \texttt{torch.utils.data.DataLoader} class, a Python iterable over a dataset, aimed to enable efficient data loading and preprocessing. The data loader supports map-style and iterable-style datasets. Map-style datasets are commonly used when the entire dataset can be indexed efficiently (i.e., the dataset fits into node local memory). In contrast, iterable-style datasets enable the streaming of large datasets where random reads are expensive or when the batch size depends on the fetched data. The user defines the data loading order when using an iterable-style dataset, allowing for implementation optimizations such as chunk reading and dynamic batch sizes. The data loader uses a single thread because true parallelism amongst threads is disabled by Python's Global Interpreter Lock (GIL). 

To allow asynchronicity, the data loader accepts a \texttt{num\_workers} parameter to perform multi-process data loading. At each iteration, worker processes are spawned to fetch the data and then destroyed at the end of the iteration or when garbage collection occurs. The data loader also has a \texttt{pin\_memory} parameter, allowing GPUs to access data faster because it is in page-locked memory~\cite{profiling-and-improving-the-pytorch-dataloader-for-high-latency-storage}. 

PyTorch uses the Python $pickle$ utility, a popular serialization tool, to save and load the model through the \texttt{torch.save()} and \texttt{torch.load()} methods. PyTorch also has recently (October $4$, $2023$ in version $2.1$) added distributed checkpointing through the \texttt{torch.distributed.checkpoint} API. This allows the loading and saving of models to be done by multiple parallel ranks. It produces at least one file per rank and is capable of checkpointing and loading models in topologies of differing world sizes. To allow the distributed checkpoint to work on different topologies, the model checkpoint must be resharded. There is no built-in functionality for asynchronous checkpointing; if necessary, the user must implement it.


\subsubsection{TensorFlow} \label{subsection:tensorflow_api} \ TensorFlow~\cite{tensorflow-docs} offers the \texttt{tf.data} API~\cite{tf-data-api} to build data loading and preprocessing pipelines. It aims to enable the construction of complex input pipelines from simple reusable pieces. The \texttt{tf.data.Dataset} class is used to represent a sequence of samples. It can be constructed from data stored in memory or from one or more files. If the dataset can fit into node local memory, the \texttt{Dataset} object can be converted to \texttt{tf.Tensor} objects, which can then be fed to the model. Otherwise, the dataset acts as a stream, and data is loaded as needed. The \texttt{tf.data.Dataset.cache()} transformation method can be used to cache the entire dataset in memory or in local storage. However, this method should not be used with large datasets that do not fit in memory or local storage as it will lead to cache thrashing. 

Asynchronous data loading is possible by prefetching future samples and can be enabled through the \texttt{Dataset} class's \texttt{prefetch()} transformation method. The number of samples can be specified as a parameter, or users can set the parameter to \texttt{tf.data.AUTOTUNE}, causing the runtime to tune the number of samples to fetch dynamically at runtime. Parallel loading of multiple datasets is possible through the dataset classes \texttt{interleave()} transformation method. Multiple parameters can be used to configure the interleave method, such as the number of datasets to read specified by the \texttt{cycle\_length} parameter and the level of parallelism to use through the \texttt{num\_parallel\_calls} parameter, which can be set to \texttt{tf.data.AUTOTUNE}, allowing TensorFlow to adjust the value at runtime dynamically. To parallelize data preprocessing, the \texttt{Dataset} object has a \texttt{map()} transformation method that takes a lambda, which can then be run across multiple CPU cores concurrently. The \texttt{map()} method takes a parameter named \texttt{num\_parallel\_calls}, which can be used to specify the level of parallelism to use. TensorFlow recommends setting this value to the number of cores available as this allows the preprocessing of samples to be run in parallel on multiple cores. If set to one, the preprocessing of samples occurs synchronously which can negatively impact training performance. In addition, users can set the value to \texttt{tf.data.AUTOTUNE} allowing TensorFlow to adjust the value dynamically at runtime.

TensorFlow has four checkpointing formats: the \texttt{SavedModel} format, \texttt{HDF5} format, \texttt{.keras} format, and \texttt{checkpoint} format. The \texttt{SavedModel} format stores the entire model graph in a binary file that is then stored within a directory containing a TensorFlow checkpoint. The format is useful for model sharing and deployment with other platforms such as TFLite~\cite{tensorflow-lite}, TensorFlow.js~\cite{tensorflow-js}, and TensorFlowHub~\cite{tensorflow-hub}. The \texttt{HDF5} format uses a single HDF5 file and is the legacy choice by TensorFlow when using the higher level \texttt{tf.keras} API. The currently suggested format is the \texttt{.keras} format. It aims to have an efficient name-based saving scheme that enables easier debugging than the legacy (i.e., HDF5) formats. The \texttt{checkpoint} format uses three files: an index file, a metadata file, and a data file. The index file contains an index of the model data, enabling fast retrieval of model values. The metadata file contains the TensorFlow computation graph, which includes the model architecture required to reconstruct the model and the data file contains the model's actual data (model weights, optimizer states, etc.). It shards the model across multiple files to enable parallelism. Asynchronous checkpointing can be enabled through \texttt{tf.train.CheckpointOptions()}. This method takes an \texttt{enable\_async} parameter, which, when set to \texttt{True}, moves the checkpointing of the model off the main thread. The return value can be passed as a parameter to the \texttt{tf.train.Checkpoint.restore()} and \texttt{tf.train.Checkpoint.save()} methods, which are used to load and save the model.

When performing distributed data parallel training, if the online dataset preparation takes more time than the model training (i.e., time spent performing forward propagation, backpropagation, and sharing of model weights), significant time can be spent waiting for samples to be preprocessed. TensorFlow offers an experimental feature named~\texttt{tf.data.experimental.service}~\cite{tf-data-service-api}, which disaggregates the online dataset preparation and training workers. This allows for the ratio of preprocessing workers to training workers to be scaled independently, ensuring that training workers do not have to wait for samples to be preprocessed. The number of online dataset preparation workers and training workers and their CPU/RAM resources can be set either manually or auto-scaled by the service. The service is also capable of sharing preprocessed data between jobs as well as executing coordinated reads to avoid stragglers, which are often caused by the varying input data sizes workers use.


\subsubsection{Scikit-learn} \ Scikit-learn~\cite{scikit-learn-docs} requires workloads with datasets that fit into node-local memory because it does not offer data loading utilities such as PyTorch's \texttt{torch.utils.data.DataLoader} or TensorFlow's \texttt{tf.data}. Instead, developers often use a separate library named Dask-ML~\cite{dask-docs} to enable the loading of large datasets. This allows models to be trained against datasets that do not fit in node-local memory. The \texttt{Client} class is the primary way of accessing large datasets in parallel. The \texttt{Client} class acts as an interface for workers to submit tasks (i.e., Python \texttt{Futures}). The Dask-ML scheduler will then schedule tasks and run them in parallel if possible. 

The \texttt{Client} class takes a parameter named \texttt{asynchronous}, which specifies whether the client acts synchronously or asynchronously. The \texttt{dask.dataframe} class has several methods that can be used to load large datasets such as \texttt{read\_parquet()}, \texttt{read\_json()}, and \texttt{read\_hdf5()}, among others. If the requested result of a task is too large to fit into node-local memory, Dask-ML chunks the result and distributes it among the memory of multiple nodes. This allows tasks submitted to the \texttt{Client} to be run in parallel by each worker that contains a partition of a distributed result. For distributed data processing, Dask-ML offers methods such as \texttt{map()} and \texttt{map\_partitions()}, which can be used to apply computations to elements or groups of elements in a collection (i.e., arrays, dataframes, and bags). In addition, reduction methods such as \texttt{mean}, \texttt{max}, and \texttt{min} are available and can be applied across partitions. 

Scikit-learn model saving and loading is commonly done through the Python $pickle$ library. Scikit-learn also recommends using the Python library $joblib$ to save objects with large $numpy$ arrays, which is common for fitted Scikit-learn estimators (i.e., any object that learns from data). Similarly to PyTorch, no built-in functionality for asynchronous checkpointing exists, and if necessary, the user must implement it.

\subsubsection{Direct NVIDIA GPU Loading~\cite{nvidia-gpudirect}} \ GPUDirect Storage~\cite{gpudirect-storage-a-direct-path-between-storage-and-gpu-memory} enables direct data loading from storage to GPU memory, and GPUDirect RDMA~\cite{gpudirect-rdma-a-direct-path-between-storage-and-network-memory} enables direct data loading from a network device to GPU memory. This avoids loading the data into the CPU memory first, preventing the creation of a bounce buffer. This saves CPU memory and reduces the overall communication overhead between devices.

\subsubsection{TensorFlow/PyTorch with Horovod~\cite{horovod}} \ Horovod is a popular distributed deep learning framework for TensorFlow, PyTorch, and Apache MXNet~\cite{mxnet}. Uber developed Horovod with the goal of making it easier for training scripts written in common ML frameworks to scale to many GPUs. It uses data parallelism to partition the dataset among the workers to enable efficient training.

\subsection{Proposed I/O Optimization Techniques}

In this section, we present the current research towards optimizing ML I/O on HPC systems. We explore a range of strategies aimed at enhancing the efficiency and scalability of ML workflows. The research presented reflects the ongoing efforts to address the unique storage challenges posed by ML.


\subsubsection*{Caching}
\ During SGD, samples are commonly shuffled based on a seed using a Pseudo Random Number Generator (PRNG). Each worker is then assigned a partition of the randomly shuffled samples and fetch batches of samples from their assigned dataset partition. The seed can be used to determine the order and frequency at which a given worker will require samples. Dryden et al.~\cite{clairvoyant-prefetching-for-distributed-machine-learning-i-o} propose NoPFS, a machine learning I/O middleware, which capitalizes on these insights to prefetch and cache samples optimally. At the application level, NoPFS can be integrated into the training pipeline of most ML frameworks with minimal code change. NoPFS reduced I/O and improved overall training time by $5.4\times$ on models using the ImageNet-1k~\cite{imagenet-a-large-scale-heirarchical-image-database}, ImageNet-22k, and CosmosFlow~\cite{cosmoflow-using-deep-learning-to-learn-the-universe-at-scale} datasets.

Importance sampling is a DNN acceleration method in which samples are skipped based on their ``importance'' value. The importance value is based on the influence a given sample has on the model's accuracy. Chen et al.~\cite{icache} introduce iCache, a novel caching approach where samples are placed into one of two in-memory caches: the H-cache or the L-cache. The placement of a sample into either cache is determined by its level of importance, with high-importance samples placed in the H-cache and low-importance samples placed in the L-cache. The technique exploits the fact that when using importance sampling, samples with higher importance values should be fed to the model more frequently than samples with lower importance values. The intelligent caching of samples based on their importance values allows I/O operations to reach the underlying storage system less frequently. 


\subsubsection*{Scheduling}
\ When using data parallel training, the reading and writing of many small files can cause cross-application I/O interference~\cite{leveraging-burst-buffer-coordination-to-prevent-i-o-interference} in the underlying PFS. Burst buffers are a reliable technique to absorb large amounts of reading and writing in a short period of time (bursts of I/O); however, when many concurrent applications are performing these operations, congestion may still occur~\cite{scheduling-the-i-o-of-hpc-applications-under-congestion}. In addition, not all HPC systems deploy burst buffers, causing scientists to look for alternative solutions. One possible technique to mitigate the congestion is using specialized I/O scheduling algorithms. Adaptive Periodic I/O scheduler (APIO)~\cite{adpatively-periodic-io-scheduling-for-concurrent-hpc-applications} is a low-profile online I/O scheduler that dynamically adjusts to the periodic and stochastic nature of ML I/O workloads to schedule I/O operations effectively. By controlling the ordering and timing of concurrent processes issuing I/O requests, APIO can mitigate potential congestion.

Due to many concurrent processes performing I/O operations during ML training, it can be difficult for middleware to characterize the status of the underlying storage nodes. This makes it challenging to schedule I/O requests to run optimally. Wang et al.~\cite{io-performance-characterization-and-prediction-through-machine-learning-on-hpc-systems} developed a lightweight ML model to predict the status of the underlying storage nodes and present a case study leveraging their model to effectively balance I/O traffic among the OSSs on a Lustre file system. They tested several models, the most accurate was a support vector machine, which achieved an accuracy of $75\%$ when predicting whether a given OSS was busy or idle.


\subsubsection*{Shuffling}
\ Various shuffling techniques have been proposed to shuffle samples between workers before each epoch. Global shuffling involves shuffling all samples across all workers, which can make caching and other I/O optimizations difficult. In contrast, local shuffling involves workers being assigned a subset of the dataset, from which they then reuse and reshuffle before each epoch. Hash-based shuffling involves applying a hash function to the keys of the samples, from which the result determines the partition to which the record will be shuffled. Apache Spark uses this technique as a primary shuffling algorithm as it can help achieve load-balancing across partitions and minimize data movement. The randomness of the shuffling does affect the convergence rate of the model~\cite{convergence-analysis-of-distributed-stochastic-gradient-descent-with-shuffling}; however, using alternative techniques may allow workers to cache the samples because they no longer read the entire dataset. Partial data shuffling, proposed by Nguyen et al.~\cite{why-globally-re-shuffle-revisiting-data-shuffling-in-large-scale-deep-learning}, achieves similar accuracy to global data shuffling while only requiring each worker to store up to $0.03\%$ of the dataset. The partial data shuffling algorithm combines the technique of global and local data shuffling. Their paper also discusses that local shuffling achieves accuracy similar to the more common global shuffling strategy in their experimentation. Integrating their solution into PyTorch requires minor modifications of the training script and a load handler of the dataset. Jie et al.~\cite{asrdataset-a-multi-granularity-shuffle-system-for-preparing-large-scale-asr-training-data} propose ASRDataset, a system aiming to speed up the training of automatic speech recognition (ASR) models by implementing chunk-level and batch-level shuffling. Their system allows the user to specify how many chunks to batch and the number of samples in each chunk. In addition, ASRDataset uses a high-speed audio data processing library implemented in C++ and CUDA, speeding up audio data augmentation and extraction. Lee and Bahn~\cite{analyzing-data-reference-characteristics-of-deep-learning-workloads-for-improving-buffer-cache-performance} proposed shuffling bundles of samples among workers. This groups samples together to increase locality. To improve buffer caching performance, in each epoch, bundles of samples are accessed in the reverse order of the previous epoch. This improves locality from the buffer cache, allowing the last bundle in the previous epoch to be the first in the next epoch. Liu et al.~\cite{efficient-data-loading-for-deep-neural-network-training} also propose block shuffling by compacting small files into blocks. They use the ImageNet dataset to test their data-loading pipeline. They cache partially loaded data to avoid redundant reads to the underlying storage system. In the first epoch, all data is cached in memory. All remaining epochs reuse half the loaded data and load the other half from disk to form batches, reducing the underlying storage accesses by half.


\subsubsection*{Asynchronicity}
\ Asynchronous I/O enables the overlapping of compute and storage. Sunwoo Lee et al.~\cite{asynchronous-io-strategy-for-large-scale-deep-learning-applications} propose an asynchronous I/O strategy to speed up data parallel model training by creating a dedicated I/O thread to run alongside each process. Their model assigns groups of the dataset to training processes and globally shuffles these groups at each epoch. In addition, each process allocates a buffer large enough to store two or more data groups. Each process's I/O thread reads groups of samples as needed into the buffer, which can occur parallel to model training. The strategy has two drawbacks. The first is the lower degree of randomness due to groups of samples being shuffled, and the second is the increased memory footprint needed to load the additional data groups asynchronously. Serizawa et al.~\cite{accelerating-machine-learning-io-by-overlapping-data-staging-and-mini-batch-generations} also propose overlapping the copying and reading of the dataset for training. They observed $1.38$ to $6.19$ times speed up when using their asynchronous strategy compared to using Lustre directly.


\subsubsection*{Compression}
\ Compression is a common technique to reduce the I/O needed to load and store data. Kuchnik et al.~\cite{progressive-compressed-records-taking-a-byte-out-of-deep-learning-data} propose using progressive compression techniques to reduce the overhead of loading data. They present a file format named progressive compressed records (PCRs), which uses progressive compression to store image datasets. Progressive compression stores data samples (images in this case) in a series of deltas, each with more fidelity. When using formats such as JPEG, data fidelity must be chosen at encoding time, often causing image datasets to be stored multiple times at varying compression levels. Using PCR, various models can be trained using the same dataset and accessed at different fidelities according to their needs. They found that PCR reduced the sizes of images by $5\%$ compared to storing the images with TFRecords. One downside to PCR is that the decoding time can be up to $2$x greater than JPEG.


\subsubsection*{Dataset distribution}
\ Many supercomputing facilities are not yet equipped with large node-local memory devices, which disables the ability to cache datasets in higher storage. Choi et al.~\cite{ddstore-distributed-data-store-for-scalable-training-of-graph-neural-networks} aim to enhance I/O performance by creating a DDStore, an in-memory data store. DDStore evenly distributes the chunks in the memory of compute nodes, after which sample accesses from a particular node are made through in-memory read transactions. To reduce possible congestion due to multiple nodes requesting a batch from the same node, DDStore replicates the dataset and distributes it among groups of workers. This allows the samples to be globally shuffled while reducing communication congestion. This strategy makes a large trade-off between reduced node-local memory and fewer PFS I/O operations. Dantas et al.~\cite{accelerating-deep-learning-training-on-hpc-systems} created Monarch, which aims to exploit storage tiering to enable fast training speeds on single-node workloads. Their solution aims to be ML framework agnostic acting as a transparent middleware which intelligently caches at different storage layers that may not fit into host memory. Aizman et al.~\cite{high-performance-io-for-large-scale-deep-learning} created a storage system named AIStore aiming to provide an infinitely scalable namespace. The namespace can be used on an arbitrary number of disks. Metadata overhead is minimized by enabling dataflow directly between compute nodes and storage targets.


\subsubsection*{Data redundancy mitigation}
\ Multiple executing workers can often prepare the same sample redundantly during the training phase. Xie et al.~\cite{a-deep-learning-dataloader-with-shared-data-preparation} introduce JOADER, a data loader aimed at reducing the redundant I/O and data preprocessing of samples. JOADER is capable of sharing data preparation work on overlapping datasets. It accomplishes this by dividing the dataset into intersection and difference sets, correlating the selections, and sharing operations that operate on the intersecting set. They also introduce a domain-specific cache policy with a novel tree data structure aimed at effectively caching samples. This strategy is more effective when the preparation of a sample takes a significant amount of time (e.g., image decoding). Jin et al.~\cite{mmdataloader} created MMDataLoader, a data preprocessing pipeline framework aiming to increase preprocessing throughput and reduce redundant data preparation. It operates at the server level, allowing it to share results across training tasks. The framework enables batches of data to be preprocessed for multiple models at the same time, eliminating the need for each task to recompute batches individually.



\subsubsection*{Storing large models}
\ With the increasing complexity of ML models, several methods have been proposed to enable the training of models that do not fit into accelerator or even clustered accelerator memory. Previous methods aiming to reduce memory usage during training include materialization and offloading. Materialization removes some activations during the forward pass. The removed activations are then recomputed during the backward pass. This strategy trades a larger computational cost for less memory. Offloading moves activations from the limited GPU memory to CPU memory during the forward pass trading data movement for memory~\cite{weight-offloading-strategies-for-training-large-dnn-models}. The limited HPC bandwidth, in addition to the increased communication caused by the offloading of the model to storage, is a large challenge for these strategies. Rajbhandari et al.~\cite{zero-infinity-breaking-the-GPU-memory-wall-for-extreme-scale-deep-learning}, propose ZeRO-Infinity to train large models by leveraging GPU, CPU, and NVMe memory. ZeRO-Infinity is fully integrated with PyTorch. Jang et al.~\cite{smart-infinity-fast-large-language-mode-training-using-near-storage-processing-on-a-real-system} propose Smart-Infinity to train large models with storage offloading. Smart-Infinity is also integrated with PyTorch and takes advantage of storage devices to form an extended memory hierarchy. In contrast to ZeRO-Infinity, they use near-storage accelerators to reduce I/O pressure and increase training speeds. In addition, gradient compression/decompression is used to reduce the storage congestion incurred from model offloading. Sun et al.~\cite{stronghold-fast-and-affordable-billion-scale-deep-learning-model-training} propose STRONGHOLD for large model training. They offload model training from the GPU to the CPU; however, the minimum amount of data to be kept on the GPU to minimize GPU memory usage is dynamically determined at runtime. In addition, they use prefetching to load layers of the model in parallel with GPU computation.

\subsubsection*{GPU storage hierarchy}
\ In GPU architectures, storage is typically organized into tiers, comprising high-bandwidth memory (HBM) and on-chip SRAM, facilitating efficient data access and processing~\cite{performance-evaluation-and-optimization-of-hbm-enabled-gpu-for-data-intensive-applications}. Doa et al.~\cite{flash-attention-fast-and-memory-efficient-exact-attention-io-awareness} propose FlashAttention which uses an I/O aware exact attention algorithm to reduce the number of memory read and writes that occur between HBM and SRAM on GPU devices. It efficiently moves data between the GPU storage tiers, resulting in faster training speeds while maintaining the same level of accuracy as traditional attention algorithms. FlashAttention has been adopted into several machine learning frameworks such as PyTorch's \texttt{nn.Transformer}, Microsoft's DeepSpeed~\cite{microsoft-deep-speed}, and NVIDIA's MEgatron-LM~\cite{megatron-lm}. In addition, it yields the fastest in MLPerf~\cite{ml-perf} BERT training on cloud instances in MLPerf training $2.0$ (June 2022) and MLPerf training $2.1$ (November 2022).


\subsubsection*{Multi-modal pipeline overhaul}
\ FFCV~\cite{ffcv} is a library aiming to speed up the data loading and processing pipeline while easily integrating into existing PyTorch applications. FFCV offers a custom file format that can store arbitrary data modalities and enable non-sequential data reading. It stores all samples in a single file but splits the dataset into pages, which trades space for more efficient data reads. If the dataset can fit into memory, it will be cached, and every read of a sample after the first will be read from memory instead of the underlying storage system. In contrast, if the dataset does not fit into memory, FFCV enables efficient data loading by using the PRNG to prefetch data samples. In addition, it uses quasi-random data shuffling in which a permutation of the data samples is loaded into a memory buffer large enough for $batch\_size$ pages of the dataset. The model then selects from batches existing within the buffer. This reduces the number of accesses to the underlying storage system. FFCV aims to eliminate unnecessary memory allocations by having operations in the data loading and processing pipeline declare memory requirements so that memory allocation occurs only once before each epoch. It also uses just-in-time (JIT) compilation, where possible, during the data processing pipeline to convert Python code to machine code. Since the machine code is no longer under the constraints of the Python interpreter, it enables multi-threading, which by extension enables collaborative reading and writing to and from memory instead of expensive primitives, i.e., threads can share batches of data. Threads also enable data preparation (data copying and augmentations) to be run in parallel as they share the same CUDA context.

\subsubsection*{Image pipeline overhaul}
\ DIESEL+~\cite{diesel-plus-accelerating-deep-learning-tasks-on-image-datasets} is an image data loading and preprocessing pipeline that can be integrated with frameworks such as TensorFlow and PyTorch. The pipeline aims to speed up the training of models that use JPG datasets. Diesel+ has a \textit{Fast Mode} in which JPG files are first processed into an intermediate \textit{fast-binary} format and stored in memory for future usage, enabling reduced loading times. There is also a \textit{Compatible Mode} for images that require no manipulation. The time required to load JPG images is further reduced by applying an online region of interest (ROI) technique to decode the images. JPG images are combined into chunks to mitigate the small random I/O reads often performed during training. These chunks are then shuffled amongst the workers, from which random individual samples can be selected in memory. In addition, a per-task distributed cache is employed across the workers to reduce the number of reads reaching the underlying storage system. Finally, to minimize metadata overhead, tasks take metadata snapshots, eliminating the contention over the metadata server.

\subsubsection*{Remote storage}
\ Deep Lake~\cite{deep-lake-a-lakehouse-for-deep-learning} is a data lake for deep learning. It can be used with object storage systems such as AWS S3 \cite{aws-s3}, Google Cloud Storage (GCS)~\cite{google-cloud-storage}, POSIX file systems, or local in-memory storage. Deep Lake has a Tensor Query Language that extends the SQL Parser from Hyrise \cite{hyrise}. This enables multidimensional array operations and deeper integration with the data lake, including indexing, slicing of arrays, and a large set of convenience functions. There is also a data visualizer made available through a web interface, allowing for the visualization of large-scale data. The Deep Lake streaming data loader delegates data fetching to C++ processes to avoid Python's GIL. Deep Lake also has a version control system that keeps track of different branches of the dataset in the same storage. This allows queries to be written which specify the dataset version, allowing for more granular and depthful searches. Deep Lake also is capable of storing non-structured data and metadata in a columnar format resulting in fast data streaming speeds.


\subsubsection*{Checkpointing}
\ With the growing size of AI models, efficient checkpointing methods are needed to ensure scalability. Nicolae et al.~\cite{deepfreeze-towards-scalable-asynchronous-checkpointing-of-deep-learning-models} propose DeepFreeze, an asynchronous checkpointing technique built using Very Low Overhead Checkpoint-Restart (Veloc)~\cite{veloc-towards-high-performance-adaptive-asynchronous-checkpointing-at-large-scale}. Veloc is a lightweight concurrency-optimized checkpointing library aimed at delivering high-performance checkpointing on top of heterogeneous storage hierarchies. DeepFreeze is a transparent checkpointing solution for synchronous data parallel training with layer-wise model parallelism. To increase checkpointing speeds, an asynchronous multilevel checkpointing approach is used in which a local copy of the learning model is captured and persisted in lightweight (i.e., node-local storage of neighboring nodes) and heavy (i.e., PFS) storage to block training for the least amount of time possible. To reduce the overhead of model serialization, a compact binary format is used that leaves out details such as tensor labels. The model's layers are sliced into shards equal to the number of MPI ranks. The ranks then write checkpoints for their partition of the model to local storage. This allows for models to be checkpointed in parallel, avoiding the stragglers problem. Anthony et al.~\cite{evaluating-multi-level-checkpointing-for-distributed-deep-neural-network-training} also propose local process model snapshots, which are then stored in node-local storage. Their method uses the SCR-Exa library~\cite{scr-exa-ehanced-scalable-checkpoint-restart-library-for-next-generation-exascale-computing}, which is a scalable checkpointing tool for HPC applications. There are two restart methods (cold and hot) available. A cold restart attempts to restart the application using the checkpointing cache within the same allocation of nodes. The hot restart tries to replace faulty nodes with idle spare nodes. Check-N-Run~\cite{check-n-run-a-checkpointing-for-training-deep-learning-recommendation-models} is a checkpointing system for deep-learning recommendation models developed by Facebook. It uses two techniques to address size and bandwidth: differential checkpointing, where only a fraction of the model changes per iteration, and quantization techniques to reduce checkpoint size. The checkpoint system is implemented under the PyTorch framework. Check-N-Run keeps training accuracy as a priority as they state an accuracy loss as small as $0.01\%$ is unacceptable. Maurya et al.~\cite{datastates-llm} introduce DataStates-LLM, an asynchronous multi-level checkpointing approach to speed up LLM checkpointing. By exploiting the periods of time when tensors and optimizer state shards are immutable, the strategy lazily copies shards of the model from the GPU during periods of immutability into a single host and streams the partial checkpoints asynchronously to persistent storage. The copying from GPU to host and the transfer from host to persistent storage can occur in parallel.

\begin{tcolorbox}[colback=white!5!white,colframe=black!75!black,title=Summary $\#5$]
Modern ML frameworks offer many optimization techniques to enable fast training speeds. There have been many recent advances aiming to optimize ML I/O in HPC systems. From optimized batch shuffling techniques to specialized I/O schedulers, I/O remains a prominent and continually evolving area of research and development. 
\end{tcolorbox}

\section{Future Research Directions of I/O for ML}
\label{sec:gaps-in-ml-io-research}



While the ML surge is rushing forward driven by rapidly increasing data collection and application use cases, data management and I/O software have been significantly lagging, especially in the HPC domain. In this section, we identify numerous data management and I/O areas that need to be improved to ensure continued progress and innovation in the training and deployment of ML models.

\subsubsection*{I/O characterization and benchmarking} \ While the DLIO benchmarks suite, which is used as the MLPerf Storage Benchmark Suite,  focuses on deep learning applications, many more AI models exist and have been evolving whose I/O patterns have not been represented in existing benchmarks. Some significant examples include reinforcement learning, Q-learning, and Deep Q Networks, widely employed for decision-making in dynamic environments. These ML models may produce unique I/O patterns, motivating the need for additional benchmarks that can mimic these patterns. Benchmarks that can properly emulate the vast number of AI models play a crucial role in assisting researchers in identifying optimization techniques, ensuring that I/O does not become a bottleneck in ML workloads. Development of I/O benchmarks representing ML used specifically in science applications that run on HPC systems is also needed.

\subsubsection*{Efficient data preparation strategies} \ When preparing data from remote storage locations such as data warehouses, it may be necessary to store intermediate forms of the dataset after preprocessing. Further research could involve an I/O middleware that transparently stores intermediate forms of the dataset in clustered memory or the underlying storage system, depending on the dataset size. By extending the memory to the underlying storage system, additional questions arise, such as how data movement between memory and storage can be optimized for high throughput and low latency and what caching strategies should be used to ensure efficient data access.

\subsubsection*{Data quality evaluation and monitoring methods} \ With the increasing amount of data for machine learning, ensuring data quality is essential for model accuracy~\cite{overview-and-importance-of-data-quality-for-machine-learning-tasks}. A step towards ensuring high-quality data is to use data warehouses because they often provide data governance tools. However, scientists who use HPC systems for high I/O throughput could take a significant performance hit if limited by the bandwidth of the external network. This motivates the need for tools that can aid in obtaining and monitoring high-quality datasets. These tools could offer data management services, such as integrity constraints, versioning, provenance, and quality assessments, while being scalable and supporting multiple data modalities. A survey on metrics for AI data readiness ~\cite{hiniduma2024data} exposes a lack of metrics focusing on quality, bias, and FAIR (Findability, Availability, Interoperability, and Reusability) principle compliance.

\subsubsection*{Support for multi-modal data} \ Multi-modal machine learning models aim to process and relate information from multiple data modalities \cite{multimodal-machine-learning-a-survey-and-taxonomy}. I/O optimization strategies targeting multi-modal datasets are needed to ensure reasonable training and inference speeds. Two broad categories for the relationship between the datasets are parallel and non-parallel data. In parallel data, each sample in one modality corresponds to another in a separate modality. In contrast, in non-parallel data, the samples within each modality are independent. Future research could exploit data locality by ensuring optimal data layout grouping related data together depending on whether it is a parallel or non-parallel dataset.

\subsubsection*{Support for streaming data} \ Due to the rise of cloud-based storage being used by ML workloads, methods of efficiently streaming data from multiple data sources are needed. In addition, multi-modal models may need to stream data from several data sources, each with varying data modalities. Future research could involve an I/O middleware capable of concurrently optimizing I/O from multiple data sources. The middleware would take into account the varying latencies between the data sources and use methods such as sample caching and sample prefetching to ensure optimal performance. This would involve a cache eviction strategy in which eviction is based on the dynamic latency of the data sources. In addition, prefetch scheduling could also occur dynamically, ensuring that samples from high-latency data sources have a higher priority than low-latency data sources. It would also take into account the sample size when performing these optimizations to ensure the optimal balance between latency and data size.

\subsubsection*{ML application specific I/O optimizations on (1) traditional HPC systems, and (2) heterogeneous systems with GPUs and domain-specific processing units} \ GPUs have become increasingly popular due to the rise of ML. This is due to their high level of parallelism when doing mathematical computations and, in particular, matrix operations~\cite{gpus-for-machine-learning}. This is possible because GPUs, unlike CPUs, contain thousands of cores. This has led to the development of ML domain-specific processing units, such as tensor processing units (TPU)s by Google~\cite{google-tpu}. These are specialized processors specifically built for ML workloads. With the wide range of hardware available, future research is needed to determine how data movement and storage techniques should be used effectively across the diverse range of hardware available. Questions for future research include where to place and process samples to prioritize processing units and exploit their contrasting abilities effectively, how to efficiently move data between the diverse range of processing units, and how to provide transparent and reasonable solutions for developers facing these challenges.

\subsubsection*{HPC I/O inference analysis} \ The large majority of current ML I/O research is aimed at analyzing the training of ML models. However, the I/O that occurs during model deployment differs widely from training because there is no longer a need to select random samples of data. This causes many of the previously mentioned I/O optimizations in Section~\ref{sec:ml-io-optimization-techniques} to be non-applicable. A thorough analysis of the I/O operations that occur during model deployment should be conducted in order to identify potential I/O bottlenecks. The study could examine a range of model deployment scenarios, including single-model workloads versus multiple-model workloads, as well as different types of models. In addition, while many models no longer need checkpointing, investigation of I/O write operations that occur due to the need to store model predictions could produce valuable insight.

\subsubsection*{Integration of HPC I/O and ML programming tools} \ As shown in Fig. \ref{fig:hpc-vs-dl-io-stack}, HPC I/O and ML I/O stacks have been organically evolving for faster implementation of the ML algorithms although there is significant scope for higher performance. As the ML algorithms settle, improving the efficiency of data access in running various applications both on HPC systems as well as cloud computing environments need to take priority. ML application development should also target obtaining portability and efficiency on different platforms. Achieving this will require efforts in benchmarking, evaluation of performance with various programming tools, and developing platform-specific optimization strategies. 

\subsubsection*{Efficient Model Offloading} \ Model offloading is used in memory-constrained environments to move subsets of the model to lower storage tiers. This has been a prominent area of research in edge computing, where memory constraints are a common limiting factor. With the growth in size of LLMs, many cannot be held in distributed memory. One example of such a model is the DOE's recently announced one-trillion-parameter (i.e., 4TiB) AI system named ``AuroraGPT''~\cite{science-gpt}. Enabling such a model in a memory-constrained environment requires model offloading. Compounding the issue is that some inference configurations (e.g., ensemble models or pipeline models) require multiple models to be loaded and unloaded from device memory to respond to a single query. Moving the model effectively through the storage tiers to minimize inference response times is a challenging I/O and data management problem where further R\&D is needed.

\begin{tcolorbox}[colback=white!5!white,colframe=black!75!black,title=Summary $\#6$]
The rapid development of AI and its usage in decision support has pushed for quick, easy, and often dirty ways to implement them. As the applications and technologies mature, efficiency will be a priority. We listed various gaps in current research and practice. Addressing the widening gap between computing and storage, additional R\&D in I/O and data management are essential to sustain progress and innovation and achieve efficiency in ML model training and development.
\end{tcolorbox}

\section{Conclusion}
\label{sec:conclusion}
Data plays a pivotal role in machine learning, impacting the learning process and significantly affecting model performance and accuracy. With a wide array of file formats available, each offering distinct I/O optimizations, machine learning applications must carefully select the format that best suits their needs. Additionally, the modality of the dataset influences the choice of file formats, affecting how datasets are accessed and stored throughout the ML lifecycle.

Model training is a critical step in this lifecycle, and due to SGD being the most popular optimization algorithm for training models on large datasets, high amounts of small random I/O reads present a significant challenge for PFSs. Furthermore, training distribution strategies such as model parallelism and data parallelism, in addition to checkpointing, exacerbate the number of I/O operations PFSs need to handle. In order to find I/O bottlenecks in ML processing, training, and inference, it is essential to have access to appropriate benchmarks and profiling tools. Existing benchmarks and profilers allow I/O performance enthusiasts and HPC practitioners to simulate ML workloads accurately, facilitating the identification of areas of improvement within ML pipelines.

Modern ML frameworks commonly offer several options to optimize data loading and processing, such as parallel data preparation, dataset streaming, and sample prefetching, to optimize the processing and loading of batches of samples. However, caching solutions that work with datasets that do not fit into node local memory are not widely available in ML frameworks such as PyTorch and TensorFlow, often leading to sample accesses always reaching the underlying storage system, which may lead to an I/O bottleneck. Recent advances toward minimizing I/O time include specialized shuffling techniques, I/O schedulers, and domain-specific file formats.

While the rate of ML I/O research has continued to increase alongside the popularity surge of AI, there are still gaps in existing research. All-in-one solutions to existing problems are a challenge for researchers due to the wide variety of data sources where samples are stored, the diverse range of hardware ML training and inference runs on, and the many different types of data used by ML models. An additional challenge is providing solutions to these problems that are either transparent or easily incorporated by ML developers. Ultimately, ongoing R\&D in ML I/O is a crucial process to ensure efficient model training and deployment, ultimately aiding in the expanding and ever-growing field of AI.

\begin{acks}
\noindent This research was supported by the U.S. Department of Energy (DOE), Office of Science, Office of Advanced Scientific Computing Research (ASCR) under contract number DE-AC02-05CH11231 with LBNL. This research used resources of the National Energy Research Scientific Computing Center, a DOE Office of Science User Facility supported by the Office of Science of the U.S. Department of Energy under Contract No. DE-AC02-05CH11231. 
\end{acks}

\bibliographystyle{ACM-Reference-Format}
\bibliography{interfaces,library,survey}


\end{document}